\def\editmode{0}

\def\bibfilenames{shared_refs}

\def\spsformat{0}

\if\spsformat1

\documentclass{article}
\usepackage{spconf}

\else
\documentclass[10pt,final,twocolumn]{IEEEtran}

\fi

\if\editmode1  
\usepackage[backend=bibtex,style=alphabetic,sorting=debug,maxbibnames=99]{biblatex}
\DeclareFieldFormat{labelalpha}{\thefield{entrykey}}
\DeclareFieldFormat{extraalpha}{}
\bibliography{\bibfilenames}
\newcommand{\cmt}[1]{\noindent\textcolor{lightgreen}{\underline{[#1]}}} 
\newcommand{\hc}[1]{\textcolor{blue}{#1}} 
\newenvironment{myitemize}{\begin{itemize}}{\end{itemize}}
\newcommand{\myitem}{\item}

\else
\usepackage{cite}
\newcommand{\cmt}[1]{} 
\newcommand{\hc}[1]{{#1}} 
\newenvironment{myitemize}{}{}
\newcommand{\myitem}{}

\fi

\newcommand{\printmybibliography}{\if\editmode1 
\printbibliography
\else
\bibliography{\bibfilenames}
\fi
}

\usepackage{fixltx2e}

\usepackage{graphicx}

\usepackage{subcaption}              
\usepackage{caption}

\usepackage[utf8]{inputenc}

\usepackage{amsfonts}
\usepackage{amsmath}
\usepackage{mathtools}

\usepackage{amssymb}

\usepackage{bm}
\usepackage{bbm}

\usepackage{color,verbatim}
\usepackage{multirow}
\usepackage{accents}

\usepackage{theoremref}

\usepackage{hyperref}

\newcounter{rulecounter}
\newcommand{\resetrule}{ \setcounter{rulecounter}{0}}
\resetrule

\newsavebox{\selvestebox}
\newenvironment{colbox}[1]
  {\newcommand\colboxcolor{#1}%
   \begin{lrbox}{\selvestebox}%
   \begin{minipage}{\dimexpr\columnwidth-2\fboxsep\relax}}
  {\end{minipage}\end{lrbox}%
   \begin{center}
   \colorbox{\colboxcolor}{\usebox{\selvestebox}}
   \end{center}}

\definecolor{orange}{rgb}{1,0.8,0}
\definecolor{gray}{rgb}{.9,0.9,0.9}
\definecolor{darkgray}{rgb}{.3,0.3,0.3}
\definecolor{darkblue}{rgb}{.1,0.0,0.3}
\definecolor{lightblue}{rgb}{0.7,0.7,1}
\definecolor{lightred}{rgb}{1,0.7,.7}
\definecolor{purple}{RGB}{204,153,255}
\definecolor{lightgray}{rgb}{.95,0.95,0.95}
\definecolor{lightgreen}{rgb}{0.3,0.5,0.3}
\definecolor{darkgreen}{rgb}{0.05,0.3,0.05}

\newcommand{\brackets}[1]{#1}

\newcommand{\tbm}[1]{{\tilde{\bm #1}}}

\newcommand{\inv}{^{-1}}

\newcommand{\rfield}{\mathbb{R}}

\newcommand{\trnb}{\mathop{\rm Tr}}

\newcommand{\transpose}{^\top}
 \newcommand{\define}{\triangleq}

\newcommand{\expected}{\mathop{\mathbb{E}} }
\newcommand{\cov}{\mathop{\textrm{Cov}}}

\newcommand{\var}[1]{\mathop{\textrm{Var}}\brackets{#1} }
\newcommand{\normal}{\mathcal{N}}

\newcommand{\minimize}{\mathop{\text{minimize}}}

\newtheorem{myproposition}{Proposition}
\newtheorem{myremark}{Remark}
\newtheorem{myproblemstatement}{Problem Statement}
\newtheorem{mylemma}{Lemma}
\newtheorem{mytheorem}{Theorem}
\newtheorem{mydefinition}{Definition}
\newtheorem{mycorollary}{Corollary}

\newcommand{\region}{\hc{\mathcal{X}}}
\newcommand{\regiondim}{{\hc{{m}}}}
\newcommand{\loc}{\hc{\bm x}}
\newcommand{\locmat}{\hc{\bm X}}

\newcommand{\txpow}{\hc{P}_\text{Tx}}

\newcommand{\pow}{\hc{r}}    
\newcommand{\powmeas}{\tilde{\pow}} 
\newcommand{\powest}{\hat{\pow}}    

\newcommand{\gridx}{\hc{N_x}}  
\newcommand{\gridy}{\hc{N_y}}   
\newcommand{\griddim}{\hc{\gridy \times \gridx}} 

\newcommand{\powmat}{\hc{\bm{R}}}  

\newcommand{\uncertnn}{\hc{\bm{V}}} 
\newcommand{\varmat}{\hc{\hat{\uncertnn}}}  

\newcommand{\powmeasmat}{\hc{\tilde{\bm \powmat}}}     

\newcommand{\measset}{\hc{\Omega}}  
\newcommand{\mask}{\hc{\bm{M}}}  
\newcommand{\buildingset}{\hc{\mathcal{B}}} 
\newcommand{\buildingsetcomp}{\hc{\bar{\mathcal{B}}}} 
\newcommand{\powmeaswithmask}{\hc{\check{\powmat}}} 

\newcommand{\weight}{\hc{\theta}}
\newcommand{\weightvec}{\hc{\bm{\weight}}}

\newcommand{\estweightvec}{\hc{\hat{\weightvec}}} 

\newcommand{\dataind}{\hc{d}} 
\newcommand{\datanum}{\hc{D}} 
\newcommand{\datanot}[1]{^{(#1)}} 

\newcommand{\meanoutfunc}{\hc{{f}^{\powmat}}}  
\newcommand{\stderroutfunc}{\hc{{f}^{\uncertnn}}} 
\newcommand{\meanfunc}{\hc{{f}^{\powmat}_{\weightvec}}}  
\newcommand{\varfunc}{\hc{{f}^{\uncertnn}_{\weightvec}}} 
\newcommand{\meannn}{\hc{{g}^{\powmat}_{\weightvec_{\powmat}}}}  
\newcommand{\varnn}{\hc{{g}^{\uncertnn}_{\weightvec_{\uncertnn}}}} 
\newcommand{\estmeanfunc}{\hc{{f}^{\powmat}_{\estweightvec}}} 

\newcommand{\encoder}{\hc{\bar{E}}} 
\newcommand{\encoweightvec}{\hc{\weightvec_{\encoder}}} 
\newcommand{\decoder}{\hc{\bar{D}}} 
\newcommand{\decoweightvec}{\hc{\weightvec_{\decoder}}} 
\newcommand{\latent}{\hc{\bm \zeta}} 

\newcommand{\covmat}{{\hc{\bm{C}}}} 
\newcommand{\mean}{{\hc{{\mu}}}} 
\newcommand{\meanvec}{{\hc{\bm{\mu}}}} 
\newcommand{\updatelinearvec}{{\hc{\bm{a}}}} 
\newcommand{\updatelinearoffset}{{\hc{{b}}}} 
\newcommand{\measlikelihoodvar}{{\hc{{\lambda}}}} 

\newcommand{\powvec}{\hc{\bm\pow}}
\newcommand{\gridpowvec}{\hc{\powvec}^{\grid}}
\newcommand{\gridpowvecest}{\hc{\hat \powvec}^{\grid}}

\newcommand{\fsgain}{{\hc{\bar l}}} 
\newcommand{\basepow}{{\hc{l}}} %
\newcommand{\basepowvec}{{\bm \basepow}} %
\newcommand{\shad}{{\hc{ s}}} 
\newcommand{\dist}{\hc{\delta}_{\shad}}
\newcommand{\shadvec}{{\bm \shad}} 
\newcommand{\ushad}{{\hc{\bar \shad}}} 
\newcommand{\ushadmean}{\mean_{\hc{\bar \shad}}} 
\newcommand{\ushadvar}{{\hc{\sigma^2_{\hc{ \shad}}}}} 
\newcommand{\fad}{{\hc{w}}} 
\newcommand{\fadvec}{{\bm \fad}} %
\newcommand{\fadvar}{{\hc{\sigma^2_{\fad}}}} 

\newcommand{\tind}{{\hc{t}}} 
\newcommand{\auxtind}{{\hc{\tau}}} 
\newcommand{\tnot}[1]{_{#1}} 
\newcommand{\tfun}{{\hc{T}}} 
\newcommand{\tupdate}{{\hc{t}_\text{upd}}} 

\newcommand{\measpow}{\tilde{\pow}}
\newcommand{\measpowvec}{{\tbm{\pow}}}
\newcommand{\measnoise}{{\hc{z}}}
\newcommand{\measnoisevec}{\bm{\measnoise}}
\newcommand{\measnoisevar}{{\hc{\sigma^2_{\measnoise}}}} 
\newcommand{\measandfadvar}{\hc{{\sigma}^{2}_{\fad,\measnoise}}}

\newcommand{\grid}{\hc{\mathcal{G}}}
\newcommand{\gridnum}{{\hc{{G}}}} 
\newcommand{\gridind}{{\hc{{g}}}} 
\newcommand{\gridnot}[1]{_{#1}} 
\newcommand{\gridloc}{{\loc}^{\grid}}

\newcommand{\gridbasepowvec}{\basepowvec^{\grid}} 
\newcommand{\gridshadvec}{\shadvec^{\grid}}
\newcommand{\gridfadvec}{\fadvec^{\grid}}

\newcommand{\uncert}{\hc{{\bar u}}} 
\newcommand{\navuncert}{\hc{{ u}}} 
\newcommand{\uncertsmooth}{\hc{\beta}} 

\newcommand{\costfun}{\hc{c}} 
\newcommand{\decfunc}{\hc{\phi}}  
\newcommand{\cost}{\hc{C}} 
\newcommand{\costweight}{\hc{\eta}} 

\newcommand{\lossweight}{\hc{\alpha}} 
\newcommand{\weightscaling}{\hc{\lambda}} 
\newcommand{\weightscalmat}{\hc{\bm{K}}} 
\newcommand{\lossdiff}{\hc{\Delta}}  
\newcommand{\meanloss}{\hc{\lossdiff_{\dataind, \weightvec}^{\powmat}}} 
\newcommand{\sigmaloss}{\hc{\lossdiff_{\dataind, \weightvec}^{\uncertnn}}} 

\newcommand{\matrowind}{\hc{i}} 
\newcommand{\matcolind}{\hc{j}} 
\newcommand{\matind}{\hc{\matrowind, \matcolind}} 

\newcommand{\velocity}{\hc{v}} 

\newcommand{\setindices}{\hc{\mathcal{T}}} 

\newcommand{\initind}{\hc{\text{I}}} 
\newcommand{\destind}{\hc{\text{F}}} 

\newcommand{\posconst}{\hc{\epsilon}} 

\usepackage{flushend}

\renewcommand{\paragraph}[1]{\textbf{#1.}}
\newcommand{\nextv}[1]{} 

\def \dcolformat{1} 
\if \dcolformat1
\newcommand{\dcol}[1]{#1} 
\else
\newcommand{\dcol}[1]{} 
\fi

\begin{document}

\title{Spectrum Surveying: Active Radio Map 
\\Estimation with Autonomous UAVs}

\if\spsformat1
\name{authors name\thanks{Thanks to XYZ agency for funding.}}
\address{Author Affiliation(s)}
\else
\author{Raju Shrestha$^1$, Daniel Romero$^1$, and Sundeep Prabhakar Chepuri$^2$ \\

$^1$Department of Information and Communication Technology, \\University of Agder, Norway.\\
$^2$Department of Electrical Communication Engineering,\\ Indian
Institute of Science, India. 
\thanks{Research funded by the Research Council of Norway (IKTPLUSS
grant 280835) and Department of Science and Technology, India. Emails: \{raju.shrestha, daniel.romero\}@uia.no, spchepuri@iisc.ac.in. Parts of this work were presented at the IEEE International Workshop on Machine Learning for Signal Processing 2020~\cite{romero2020surveying}.} } 

\fi

\maketitle
\begin{abstract}
\begin{myitemize}%
  \myitem\cmt{Radio maps}Radio maps find numerous applications in wireless communications and mobile robotics tasks, including resource allocation, interference
coordination, and mission planning. 
\begin{myitemize}%
  \myitem\cmt{Spectrum Cartography (SC) techniques}Although numerous
  techniques have been proposed to construct radio maps from spatially
  distributed measurements, \myitem\cmt{Limitations}%
\begin{myitemize}%
  \myitem\cmt{Fixed locations}the locations of such measurements are
  assumed predetermined beforehand.
\end{myitemize}%
\end{myitemize}%
\myitem\cmt{Proposed spectrum surveying}In contrast, this paper
proposes \emph{spectrum surveying}, where a mobile robot such as an
unmanned aerial vehicle (UAV) collects measurements at a set of
locations that are actively selected to obtain high-quality map
estimates in a short surveying time. This is performed in two steps. 
\begin{myitemize}%
  \myitem\cmt{Estimators for uncertainty mapping}First, two novel algorithms, a model-based online Bayesian estimator and a data-driven deep learning algorithm, are devised for updating a map estimate and an uncertainty metric that indicates the informativeness of  measurements at each possible location.
  These
  algorithms offer complementary benefits and feature constant
  complexity per measurement.
  \myitem\cmt{Trajectory for measurement collections}Second, the
  uncertainty metric is used to plan the trajectory of the UAV to
  gather measurements at the most informative locations. To overcome
  the combinatorial complexity of this problem, a dynamic programming
  approach is proposed to obtain lists of waypoints through areas of
  large uncertainty in linear time.  
\end{myitemize}%
\myitem\cmt{Numerical result}Numerical experiments conducted on a realistic dataset confirm that the
proposed scheme constructs accurate radio maps quickly.
\end{myitemize}%
\end{abstract}

\section{Introduction}
\cmt{Introduction}%
\begin{myitemize}%
  \myitem\cmt{overview of radio maps}Radio maps are functions that
  provide a certain channel metric, such as received signal strength,
  power spectral density, or channel gain, across a given geographical
  area. 
  \myitem\cmt{Applications of radio maps}%
\begin{myitemize}%
  \myitem\cmt{General}Radio maps are commonly used in tasks such as network
  planning, interference coordination, power control, spectrum
  management, resource allocation, hand-off procedure design, dynamic
  spectrum access, and cognitive radio~\cite{grimoud2010rem,
    yilmaz2013radio}, to name a few.
  \myitem\cmt{UAVs communication}Besides, radio maps are gaining
  popularity for autonomous unmanned aerial vehicle (UAV)
  communications in tasks such as mission
  planning~\cite{zeng2021simultaneous} or optimal relay deployment in
  UAV-assisted networks; see, e.g.~\cite{chen2017map,
    zhang2018cellular, chen2017learning, chen2019efficient}. 
\end{myitemize}%
\myitem\cmt{motivation}All these applications require methods for
accurate radio map construction.
\end{myitemize}%

\cmt{literature}%
\begin{myitemize}
  \myitem\cmt{General spectrum cartography approaches}
  A large number of approaches have emerged to address the radio map
  estimation problem. They rely on measurements acquired by spatially
  distributed sensors, possibly integrated into user equipment such as
  mobile phones, to construct radio maps by means of some form of
  spatial interpolation.
\begin{myitemize}%
\myitem\cmt{Model-based (Not based on deep learning)}%
\begin{myitemize}%
  \myitem\cmt{Power maps}Schemes to construct \emph{power} maps, which
  provide the received signal strength across space, have been
  developed using kriging~\cite{alayafeki2008cartography,
    boccolini2012wireless, agarwal2018spectrum, romero2020surveying},
  dictionary learning~\cite{kim2013dictionary}, sparse Bayesian
  learning~\cite{huang2014cooperative, yang2016compressive,
    he2018steering}, and matrix completion~\cite{hu2013efficient}.
  \myitem\cmt{PSD maps}Power spectral density (PSD) maps can be
  estimated using kernel-based learning~\cite{romero2017spectrummaps,
    bazerque2013basispursuit, teganya2019locationfree}, sparse learning
  \cite{bazerque2013basispursuit}, and tensor completion
  \cite{tang2016spectrum, zhang2020spectrum}.
  \myitem\cmt{Channel gain maps}The problem of estimating channel-gain
  maps has been addressed in~\cite{lee2018adaptive, lee2017lowrank,
    xu2021hierarchical}.
  \myitem\cmt{Limitation of model-based SC}All the aforementioned
  approaches are based on interpolation algorithms or rely on modeling
  propagation phenomena~\cite{jayawickrama2013compressive}. However, the
  actual radio propagation environment is often complex and
  inappropriate modeling will generally lead to poor estimation
  performance.
\end{myitemize}%
\end{myitemize}%
\begin{myitemize}%
  \myitem\cmt{Deep learning-based SC}To bypass this problem,
  \cite{teganya2020rme, han2020power, shrestha2021deep,
    saito2019twosteppathloss} proposed data-driven approaches where a
  deep neural network (DNN) is used to learn the underlying
  propagation phenomena such as shadowing, reflection, and diffraction
  from a dataset of measurements.
\end{myitemize}%
\myitem\cmt{Limitation of existing SC methods}%
\begin{myitemize}%

  \myitem\cmt{Not suitable for Mobile sensing}All the
  preceding methods assume that the measurement locations are given
  and, as a result, cannot decide where to measure next, which is
  necessary in certain contexts. \end{myitemize}%
\end{myitemize}%
\cmt{Proposed Work}%
\begin{myitemize}%
  \myitem\cmt{Goal}Specifically, certain applications such as network
  planning or those involving UAV communications demand approaches to
  construct radio maps by \emph{surveying} the channel at the area of
  interest with a sensor on board a mobile robot. To this end, this
  work puts forth an active sensing method where an autonomous
  vehicle, such as a UAV, equipped with a sensing and communication
  module collects measurements at a judiciously selected set of
  locations to efficiently construct a
  high-quality radio map.
  \myitem\cmt{Need of uncertainty metric}In particular, the set of
  most informative measurement locations is selected to approximately minimize the
  operation time required to attain a certain estimation
  accuracy. Towards this purpose, suitable metrics that provide the
  uncertainty associated with every spatial location are
  developed. Using the collected measurements, the UAV computes this
  metric on-the-fly and plans a trajectory accordingly.

  \myitem\cmt{Contributions}The contributions\footnote{The conference
    version of this work~\cite{romero2020surveying} presents the idea of
    the online Bayesian algorithm for model-based uncertainty mapping
    and spectrum surveying with a UAV.  Relative to
    \cite{romero2020surveying}, the present paper additionally
    proposes a data-driven uncertainty mapping scheme using a DNN, a new
    trade-off scheme for trajectory planning, and extensive
    empirical validation and comparison with existing benchmarks
    through a realistic dataset.}  of this work are as follows:
\begin{itemize}
  \item[C1)]\cmt{Contr. 1: Model-based Online uncertainty mapping}An online
  Bayesian learning scheme is proposed for joint radio map estimation
  and uncertainty mapping with constant complexity per
  measurement. This is required since UAVs process measurements as
  they become available and update their trajectory accordingly.
The proposed algorithm constitutes an online alternative to the
  popular kriging estimator~\cite{rasmussen2006gaussianprocesses}, which is grounded on the
  well-known Gudmundson shadowing model~\cite{gudmundson1991correlation}. 
  \item[C2)]\cmt{Contr. 2 :Data-driven uncertainty mapping}Complementing the
  benefits of the aforementioned online Bayesian estimator, a
  \emph{data-driven} approach for jointly estimating a power map and
  uncertainty metric is developed based on a DNN. As existing
  DNN-based radio map estimators, the proposed algorithm learns
  propagation phenomena from a dataset and, therefore, it yields a
  superior performance relative to the online Bayesian estimator at
  the expense of increasing computational complexity. 
%
  \item[C3)]\cmt{Contr. 3: Uncertainty-aware trajectory planning}A
  trajectory planning strategy is proposed for measurement acquisition
  at the most informative locations as indicated by the uncertainty
  metrics from C1) or C2).  To reduce the computational complexity of
  such a task, a simple waypoint-search approach based on the
  Bellman-Ford shortest path algorithm~\cite{bellman1958routing} is
  presented. 
\end{itemize}%
The code necessary to
reproduce all experiments is available at \url{https://github.com/uiano/spectrum_surveying_with_UAVs} and a video illustrating a spectrum surveying operation is available at \url{https://youtu.be/r9zDr4O0Fp8}.

\myitem\cmt{Novelty}The main novelty of this work is twofold: i) this paper proposes data-driven uncertainty mapping in
realistic propagation environments with possibly non-Gaussian channel distribution, and ii) this paper proposes
uncertainty-aware trajectory planning for spectrum surveying with
autonomous UAVs.

\end{myitemize}%

\cmt{Paper structure} %
\begin{myitemize}%
  \myitem\cmt{}The rest of the paper is structured as follows:
  Sec.~\ref{sec:system} formulates the problem 
  and outlines the proposed approach. Secs.~\ref{sec:model-based}
  and~\ref{sec:data-driven} respectively present model-based and
  data-driven algorithms for power map estimation and uncertainty
  mapping.  These algorithms lay the grounds for the uncertainty-aware
  trajectory planning scheme in Sec.~\ref{sec:routeplanning}. The
  proposed scheme is empirically validated in Sec.~\ref{sec:results}
  by means of data obtained from a ray-tracing simulator. Finally,
  Sec.~\ref{sec:relatedwork} draws connections with related work and
  Sec.~\ref{sec:conclusions} concludes the paper.
  
\end{myitemize}%

\cmt{Notation}%
\begin{myitemize} %
  \myitem\cmt{}\textit{Notation:}$|\mathcal{X}|$ denotes the
  cardinality of set $\mathcal{X}$. The function $\lfloor x \rfloor$
  denotes the largest integer less than or equal to $x$. Bold
  uppercase letters denote matrices or tensors, bold lowercase letters
  represent column vectors, and non-bold lowercase letters denote
  scalars. $[\bm{x}]_{i}$ is the $i$-th entry of vector $\bm{x}$,
  $\text{vec}(\bm{X})$ denotes vectorization of matrix $\bm{X}$,
  $[\bm{X}]_{i,j}$ is the $(\matind)$-th entry of matrix $\bm{X}$, and
  $[\bm{Y}]_{i,j,k}$ is the $(i,j,k)$-th entry of tensor $\bm{Y}$. The
  Hadamard product is represented by $\odot$ and $\|\bm{X}\|_{F}$
  refers to the Frobenius norm of $\bm{X}$.
\end{myitemize}

\section{Spectrum Surveying}
\label{sec:system}%
\cmt{overview}This section formulates the problem and outlines the
proposed approach.

\subsection{Problem Formulation}
\begin{myitemize}%
\myitem\cmt{Area}Let $\region \subset \rfield^{\regiondim}$ be the geographical
area of interest, where $\regiondim$ is typically $2$ or $3$, and consider $S$
transmitters, which may correspond to cellular base stations.
\myitem\cmt{Radio map}A power map is a function $\pow(\loc)$ which
provides the value of the received power, also known as signal strength,
at every $\loc\in\region$.
\myitem\cmt{Measurements} 
  A UAV equipped with a communication module capable of measuring the
  received power and a positioning system such as GPS adaptively
  chooses future measurement locations based on previous
  measurements. The power measurements and their locations are
  respectively denoted as
  $\{\measpow_{\auxtind}\}_{\auxtind=0}^{\tind}$ and
  $\{\loc_{\auxtind}\}_{\auxtind=0}^{t}$, where $\powmeas_{\auxtind} = \pow(\loc_{\auxtind}) +
  \measnoise_{\auxtind}$ and $\measnoise_{\auxtind}$ represents
  measurement noise. For convenience, the
  measurements up to $\tind+1$  will be arranged as
  $\measpowvec\tnot{\tind}\define[\measpow\tnot{0},\ldots,\measpow\tnot{t}
  ] \in \rfield^{t+1}$ and
  $\locmat_{\tind} \define[\loc\tnot{0},\ldots,\loc\tnot{t}] \in
  \rfield^{\regiondim\times(t+1)}$.
\myitem\cmt{Goal}The  goal is to determine $\locmat_{\tind}$ in order
to infer an accurate radio map as quickly as possible using the
measurements collected at these locations. 
\end{myitemize}

\subsection{Proposed Approach}
To judiciously select the future most informative locations, one needs to
solve two sub-problems:
\begin{myitemize}%
  \myitem\cmt{Radio map estimation and uncertainty
    mapping} 
  P1) Given
  $\{(\loc\tnot{\auxtind}, \powmeas\tnot{\auxtind}), \auxtind = 0,
  ..., \tind\}$, the task is to find an estimate $\powest$ of $\pow$
  and an uncertainty metric that represents the uncertainty associated
  with each location.  Function $\pow$ is typically referred to as the
  \textit{true radio map}, whereas $\powest$ is the \textit{map
    estimate}. An algorithm that produces $\powest$ is termed  a
  \textit{map estimator}.
  \myitem\cmt{Trajectory planning}P2) Using the uncertainty metric
  from P1, the second problem involves planning a trajectory for
  measurement collection to attain the desired estimation accuracy as
  fast as possible.


\end{myitemize}

\cmt{Detail formulation in later sections}The following
  sections will further detail the  formulations of these two
  problems. Secs. \ref{sec:model-based} and \ref{sec:data-driven}
  provide two algorithms with complementary benefits to address P1,
  whereas Sec.~\ref{sec:routeplanning}  is concerned with P2.


\section{Model-Based Uncertainty Mapping}
\label{sec:model-based}
\cmt{Overview}This section builds upon a widely-used probabilistic
shadowing model to develop  an \emph{online} Bayesian
algorithm for power map estimation and uncertainty mapping.
\begin{myitemize}%
  \myitem\cmt{Idea is to obtain posterior distribution}To find a
  suitable uncertainty metric, the idea is to obtain the posterior
  distribution of $\pow$ given the measurements.
  \begin{myitemize}
    \myitem\cmt{mean}While the mean of such a distribution provides the
    \emph{minimum mean square error} (MMSE) estimate of $\pow$,
    \myitem\cmt{variance}its variance can be used as uncertainty metric.  This contrasts
    with most algorithms in the literature, which just provide
    estimates of $\pow$ without any associated uncertainty metric. 
  \end{myitemize}%
\end{myitemize}%

\subsection{Radio Map Model}
\label{subsec:radiomodel}
\cmt{Single transmitter model}%
\begin{myitemize}%
\myitem\cmt{Received power}To simplify the notation,  consider a single transmitter with transmit power $\txpow$. Then, the power received at $\loc \in \region$ in logarithmic units is given by
\begin{align}
    \label{eq:powerreceived}
    \pow(\loc) = \txpow + \fsgain(\loc) - \ushad(\loc) + \fad(\loc),
\end{align}
where
\begin{myitemize}%
\myitem\cmt{Free space path loss}$\fsgain(\loc)$ quantifies free-space path loss and antenna gain, 
\myitem\cmt{Shadowing}$\ushad(\loc)$ is the loss due to shadowing, and 
\myitem\cmt{Fading}$\fad(\loc)$ is a gain due to small-scale fading and unmodeled effects.
\myitem\cmt{Shadowing and Fading Model}%
\begin{myitemize}%
\myitem\cmt{Shadowing model}The  log-normal shadowing component
$\ushad(\loc) \sim \mathcal{N}(\ushadmean, \ushadvar)$ follows the
Gudmundson correlation  model~\cite{gudmundson1991correlation}, which
prescribes that $\text{Cov}(\ushad(\loc),  \ushad(\loc^{\prime})) =
\ushadvar2^{-\|\loc - \loc^{\prime}\|/\dist}$, where $\ushadvar$ is a
constant and $\dist$ is the distance at which the correlation decays to $1/2$. 
\myitem\cmt{Fading model}Furthermore, following~\cite{huang2014cooperative}, $\fad(\loc)$ is modeled as $\mathcal{N}(0, \fadvar)$ and 
\myitem\cmt{Independence assumption}assumed independent of $\fad(\loc^{\prime})$ and $\ushad(\loc^{\prime\prime}) $ $\forall\loc^{\prime}$, $\loc^{\prime\prime} \in \region$ with $\loc^{} \neq \loc^{\prime}$. 
\end{myitemize}%
\end{myitemize}%
\myitem\cmt{Simple received power model}For brevity, rewrite~\eqref{eq:powerreceived} as
\begin{align}
    \label{eq:simplepower}
    \pow(\loc) = \basepow(\loc) - \shad(\loc) + \fad(\loc),
\end{align}
where 
\begin{myitemize}%
  \myitem\cmt{}$\basepow(\loc) \define \txpow + \fsgain(\loc) -
  \ushadmean$ and
  \myitem\cmt{}$\shad(\loc) \define \ushad(\loc) - \ushadmean$. Here
  $\basepow(\loc)$ is assumed to be a known deterministic component
  since i) $\txpow$ and the source location can be assumed known as base
  stations in contemporary cellular networks share this information
  with the users and ii) $\ushadmean$ can be estimted readily from a set
  of measurements.
\end{myitemize}%

\myitem\cmt{Grid Notation}%
\begin{myitemize}%
\myitem\cmt{Motivation}To avoid unbounded complexity for finding
estimates $\forall\loc\in\region$~\cite[Sec. 6.4]{bishop2006}, the map
and uncertainty metrics will be evaluated at a finite set of arbitrary grid points
\myitem\cmt{Grid}$\grid\define\{\gridloc\gridnot{0},\ldots,\gridloc\gridnot{\gridnum-1}\}\subset \region$.
\myitem\cmt{def}Thus, using  \eqref{eq:simplepower}, one can write
\begin{align}
\label{eq:gridpowvec}
\gridpowvec\define[\pow(\gridloc\gridnot{0}),\ldots,\pow(\gridloc\gridnot{\gridnum-1})]\transpose
= \gridbasepowvec -\gridshadvec + \gridfadvec,
\end{align}
where
\begin{myitemize}%
\myitem\cmt{}$\gridbasepowvec\define[\basepow(\gridloc\gridnot{0}),\ldots,\basepow(\gridloc\gridnot{\gridnum-1})]\transpose$, \myitem\cmt{}$\gridshadvec\define[\shad(\gridloc\gridnot{0}),\ldots,$ $\shad(\gridloc\gridnot{\gridnum-1})]\transpose$,
and
 \myitem\cmt{}$\gridfadvec\define[\fad(\gridloc\gridnot{0}),\ldots,\fad(\gridloc\gridnot{\gridnum-1})]\transpose$.
 \end{myitemize}%
\end{myitemize}

\myitem\cmt{Measurement model}The UAV collects a power measurement
\begin{myitemize}%
  \myitem\cmt{def}$\powmeas_{\auxtind} = \pow(\loc_{\auxtind}) +
  \measnoise_{\auxtind}$ when it is at position
  $\loc_{\auxtind}\in\region$, $\tau = 0, 1,\ldots$, where
  $\measnoise_{\auxtind}\sim\normal(0, \measnoisevar)$ models the
  measurement error and is assumed independent across $\auxtind$ and
  independent of $\fad(\loc)$ and $\shad(\loc^{\prime})$
  $\forall\loc$, $\loc^{\prime} \in\region$.
\end{myitemize}
\myitem\cmt{simple notation for measurements}For notational purposes, let
\begin{align}
\label{eq:measpowvec}
\measpowvec\tnot{\tind} = \basepowvec\tnot{\tind} - \shadvec\tnot{\tind} + \fadvec\tnot{\tind} + \measnoisevec\tnot{\tind},
\end{align}
where
\begin{myitemize}%
\myitem\cmt{}$\basepowvec\tnot{\tind}\define[\basepow(\loc\tnot{0}),\ldots,\basepow(\loc\tnot{\tind})]\transpose$,
\myitem\cmt{}$\shadvec\tnot{\tind}\define[\shad(\loc\tnot{0}),\ldots,\shad(\loc\tnot{\tind})]\transpose$,
\myitem\cmt{}$\fadvec\tnot{\tind}\define[\fad(\loc\tnot{0}),\ldots,\fad(\loc\tnot{\tind})]\transpose$, and
\myitem\cmt{}$\measnoisevec\tnot{\tind}\define[\measnoise\tnot{0},\ldots,\measnoise\tnot{\tind}]\transpose$.
\end{myitemize}%
\end{myitemize}%

\subsection{Batch Bayesian Estimation for Uncertainty Mapping}
\label{subsec:batch}
\cmt{Overview}To facilitate understanding, the batch version of the
problem is described before presenting the proposed online
algorithm. The batch problem, commonly referred to as
\emph{kriging}~\cite{simpson2001kriging}, 
\cmt{Batch Problem}is
\begin{myitemize}%
\myitem\cmt{formulation}%
\begin{myitemize}%
\myitem\cmt{find}to obtain the posterior distribution $p(\gridpowvec|\measpowvec\tnot{\tind},\locmat\tnot{\tind})$
\myitem\cmt{given}given $\measpowvec\tnot{\tind}$ and $\locmat\tnot{\tind}$.
\end{myitemize}%
\myitem\cmt{Analysis}%
\begin{myitemize}%
\myitem\cmt{Mean}Applying a  well-known result in estimation theory~\cite[Ch. 10]{kay1}, the mean of such a posterior distribution provides
the MMSE estimate of $\pow(\loc)$
at the grid points.
\myitem\cmt{covariance}On the other hand, the covariance of this posterior captures the uncertainty
about the true $\gridpowvec$ after observing $\measpowvec\tnot{\tind}$ and $\locmat\tnot{\tind}$.
\end{myitemize}%
\end{myitemize}%

\cmt{$\gridpowvec \perp \measpowvec\tnot{\tind}
| \gridshadvec$}
It is straightforward to establish that $\gridpowvec$ is conditionally independent of
  $\measpowvec\tnot{\tind}$ given $\gridshadvec$ using the model represented by  \eqref{eq:gridpowvec}
  and \eqref{eq:measpowvec}. As a result, it follows that 
\begin{align}
\label{eq:marginalbatch}
p(\gridpowvec|\measpowvec\tnot{\tind}) \dcol{\nonumber &}= \int p(\gridpowvec, \gridshadvec|\measpowvec\tnot{\tind})d\gridshadvec \dcol{\\ &}= \int p(\gridpowvec|\gridshadvec) p(\gridshadvec|\measpowvec\tnot{\tind})d\gridshadvec,
\end{align}
where $\locmat\tnot{\tind}$ has been omitted for brevity.
\begin{myitemize}%
\myitem\cmt{First term}From \eqref{eq:gridpowvec} and the fact that 
$\gridbasepowvec$ is deterministic, it clearly follows
that the first factor in the integrand is 
 $p(\gridpowvec|\gridshadvec)=\normal(\gridpowvec| \gridbasepowvec
 -\gridshadvec, \fadvar \bm I_\gridnum)$.
\myitem\cmt{Second term}To obtain the  second factor
$p(\gridshadvec|\measpowvec\tnot{\tind})$, observe that $\gridshadvec$
and $\measpowvec\tnot{\tind}$ are jointly Gaussian.
\begin{myitemize}%
\myitem\cmt{joint}
In particular, one
can obtain the parameters of their joint distribution
$p(\gridshadvec,\measpowvec\tnot{\tind})$ as follows. 
\begin{myitemize}%
\myitem\cmt{mean}Firstly, the mean vectors are
$\expected[\gridshadvec]=\bm 0$ and $\expected[\measpowvec\tnot{\tind}]
= \basepowvec\tnot{\tind}$. 
\myitem\cmt{covariance}Next, to compute the covariance, let $\cov[\gridshadvec]\define\covmat_{
  \gridshadvec}$ and write
 \begin{align}
 \cov[\gridshadvec,\measpowvec\tnot{\tind}]\dcol{&}=\expected[
\gridshadvec(\measpowvec\tnot{\tind} -\basepowvec\tnot{\tind}) 
\transpose]\dcol{\nonumber \\ &}
=\expected[
\gridshadvec(- \shadvec\tnot{\tind} + \fadvec\tnot{\tind} + \measnoisevec\tnot{\tind}) 
\transpose]
\dcol{\nonumber
\\&}=-\expected[
\gridshadvec \shadvec\tnot{\tind} \transpose]
 \define-\covmat_{\gridshadvec,
 \shadvec\tnot{\tind}}
 \end{align}
 as well as 
 \begin{align}
\cov[\measpowvec\tnot{\tind}]&=\expected[
(\measpowvec\tnot{\tind} -\basepowvec\tnot{\tind})
(\measpowvec\tnot{\tind} -\basepowvec\tnot{\tind})\transpose]\nonumber
\\&=\expected[
(- \shadvec\tnot{\tind} + \fadvec\tnot{\tind}
+ \measnoisevec\tnot{\tind})
(- \shadvec\tnot{\tind} + \fadvec\tnot{\tind} + \measnoisevec\tnot{\tind})\transpose]\nonumber
\\&= \cov[\shadvec\tnot{\tind}] + \fadvar\bm I_{\tind+1} 
+ \measnoisevar\bm I_{\tind+1}\dcol{\nonumber
\\&}\define \covmat_{\shadvec\tnot{\tind}} + \measandfadvar \bm I_{\tind+1}, 
 \end{align}%
 where $\measandfadvar = \fadvar\bm
+ \measnoisevar\bm$. Here, the matrices
$\covmat_{ \gridshadvec}$,
$\covmat_{\gridshadvec, \shadvec\tnot{\tind}}$ and
$\covmat_{\shadvec\tnot{\tind}}$ can be obtained from the covariance
function introduced in Sec.~\ref{subsec:radiomodel}.
\end{myitemize}%
\myitem\cmt{shadowing posterior}Applying \cite[Th. 10.2]{kay1} to this
joint distribution, it follows
that $p(\gridshadvec|\measpowvec\tnot{\tind})
= \normal(\gridshadvec|\meanvec_{\gridshadvec|\measpowvec\tnot{\tind}}, \covmat_{\gridshadvec|\measpowvec\tnot{\tind}})$, with
\begin{align}
\meanvec_{\gridshadvec|\measpowvec\tnot{\tind}}&=
\cov[\gridshadvec,\measpowvec\tnot{\tind}]{\cov}\inv[\measpowvec\tnot{\tind}](\measpowvec\tnot{\tind}-
\expected[\measpowvec\tnot{\tind}])\nonumber
\\&=
-\covmat_{\gridshadvec,
 \shadvec\tnot{\tind}}
( \covmat_{\shadvec\tnot{\tind}} + \measandfadvar \bm I_{\tind+1})\inv
(\measpowvec\tnot{\tind}- \basepowvec\tnot{\tind})
\\
\covmat_{\gridshadvec|\measpowvec\tnot{\tind}}&=
\cov[\gridshadvec]
-\cov[\gridshadvec,\measpowvec\tnot{\tind}]{\cov}\inv[\measpowvec\tnot{\tind}]
\cov[\measpowvec\tnot{\tind},\gridshadvec]\nonumber
\\&=\covmat_{\gridshadvec}
-\covmat_{\gridshadvec,
 \shadvec\tnot{\tind}}
( \covmat_{\shadvec\tnot{\tind}} + \measandfadvar \bm I_{\tind+1})\inv
\covmat_{
 \shadvec\tnot{\tind},\gridshadvec},\label{eq:batchcov}
\end{align}
where $\covmat_{ \shadvec\tnot{\tind},\gridshadvec}\define\covmat_{\gridshadvec, \shadvec\tnot{\tind}}\transpose$.
\end{myitemize}%
\end{myitemize}%
\cmt{re-composition}Finally, applying  \cite[eq. (2.115)]{bishop2006}
to obtain the conditional marginal in \eqref{eq:marginalbatch} yields
$p(\gridpowvec|\measpowvec\tnot{\tind}) = \normal(\gridpowvec|
\meanvec_{\gridpowvec|\measpowvec\tnot{\tind}}
,\covmat_{\gridpowvec|\measpowvec\tnot{\tind}} )$ with
$\meanvec_{\gridpowvec|\measpowvec\tnot{\tind}}\define
\basepowvec\tnot{\tind}-\meanvec_{\gridshadvec|\measpowvec\tnot{\tind}}$
and
$\covmat_{\gridpowvec|\measpowvec\tnot{\tind}}\define\fadvar \bm
I_\gridnum +\covmat_{\gridshadvec|\measpowvec\tnot{\tind}} $, thereby
solving the batch problem. The map estimate is, therefore,
$\gridpowvecest = \meanvec_{\gridpowvec|\measpowvec\tnot{\tind}}$
and $\covmat_{\gridpowvec|\measpowvec\tnot{\tind}}$ captures the
uncertainty in this estimate.

\subsection{Online Bayesian Estimation for Uncertainty Mapping}
\label{subsection:online}
\cmt{Online Problem}Although the solution described in
Sec.~\ref{subsec:batch} could in principle be utilized to determine
the trajectory of the UAV, it suffers from a limitation:
\begin{myitemize}%
  \myitem\cmt{Motivation}since \eqref{eq:batchcov} involves inverting
  a $(\tind+1)\times (\tind+1)$ matrix, the complexity per uncertainty
  computation will grow as more measurements become available, eventually
  becoming unaffordable. Thus, it is more convenient to adopt an
  \emph{online} approach where each new measurement is utilized to
  refine the previous posterior.
  \myitem\cmt{formulation}Specifically, consider the problem of
\begin{myitemize}%
\myitem\cmt{find}finding $p(\gridpowvec|\measpowvec\tnot{\tind},\locmat\tnot{\tind})$
\myitem\cmt{given}given the previous posterior
$p(\gridpowvec|\measpowvec\tnot{\tind-1},\locmat\tnot{\tind-1})$ and
the most recent measurement
$(\measpow\tnot{\tind}, \loc\tnot{\tind})$ with a computational
complexity that does not grow with~$\tind$.
\end{myitemize}%
\end{myitemize}%

\cmt{decomposition}
\begin{myitemize}%
\myitem\cmt{Goal}To address this problem, it is convenient to decompose
$p(\gridpowvec|\measpowvec\tnot{\tind})$ into the previous posterior $
p(\gridpowvec|\measpowvec\tnot{\tind-1})$ and a term that depends on
the current measurement but not on the previous ones. 
\myitem\cmt{conditional indep.}However, it can be easily seen
that such a factorization is not possible due to the posterior
correlation among measurements. To bypass this difficulty, the
central idea in the proposed online learning scheme is to use $\grid$ to summarize
the information of all past measurements.
\begin{myitemize}%
\myitem\cmt{mathematically}Mathematically, this can be
formulated as the assumption that $\measpow\tnot{\tind}$ and
$\measpowvec\tnot{\tind-1}$ are conditionally independent given
$\gridpowvec$. That is, when $\gridpowvec$ is known, the past
measurements $\measpowvec\tnot{\tind-1}$ do not provide extra
information about $\measpow\tnot{\tind}$. 
\myitem\cmt{assumption error reduction}The error that this
approximation introduces, which  can be reduced by adopting a denser grid,
pays off since it enables online estimation. Besides, Sec.~\ref{sec:results}
demonstrates that this error is negligible. 
\end{myitemize}%

\myitem\cmt{bayes}From this assumption and Bayes' rule,  it follows that
\begin{align}
\label{eq:bayes}
p(\gridpowvec&|\measpowvec\tnot{\tind})=p(\gridpowvec|\measpow\tnot{\tind},\measpowvec\tnot{\tind-1})\nonumber
\propto
p(\measpow\tnot{\tind},\measpowvec\tnot{\tind-1}|\gridpowvec)p(\gridpowvec)\nonumber
\\&=
p(\measpow\tnot{\tind}|\gridpowvec)p(\measpowvec\tnot{\tind-1}|\gridpowvec)p(\gridpowvec)\nonumber
=
p(\measpowvec\tnot{\tind-1},\gridpowvec)p(\measpow\tnot{\tind}|\gridpowvec)\nonumber
\\&=
p(\gridpowvec|\measpowvec\tnot{\tind-1})p(\measpowvec\tnot{\tind-1})p(\measpow\tnot{\tind}|\gridpowvec)
\propto
p(\gridpowvec|\measpowvec\tnot{\tind-1})p(\measpow\tnot{\tind}|\gridpowvec),
\end{align}
where $\propto$ denotes equality up to a  scaling constant. In
this context, a constant is understood as a positive term that does not depend
on $\gridpowvec$.
\begin{myitemize}%
  \myitem\cmt{first term}The term
  $p(\gridpowvec|\measpowvec\tnot{\tind-1})=\normal(\gridpowvec|\meanvec_{\gridpowvec|\measpowvec\tnot{\tind-1}},\covmat_{\gridpowvec|\measpowvec\tnot{\tind-1}})$
  is obtained from the previous iteration. 
\myitem\cmt{second term}To find
$p(\measpow\tnot{\tind}|\gridpowvec)$, note that $\measpow\tnot{\tind}$ 
and $\gridpowvec$ are jointly Gaussian. It follows
from \cite[Th. 10.2]{kay1} that $p(\measpow\tnot{\tind}|\gridpowvec)$
is Gaussian distributed with parameters
\begin{align}
\expected&[\measpow\tnot{\tind}|\gridpowvec]
=
\expected[\measpow\tnot{\tind}] +
\cov[\measpow\tnot{\tind},\gridpowvec]
{\cov}\inv[\gridpowvec]
(\gridpowvec-\expected[\gridpowvec])\nonumber\\
&=
\basepow(\loc\tnot{\tind}) +
\expected[(-\shad(\loc\tnot{\tind})+\fad(\loc\tnot{\tind})+\measnoise\tnot{\tind})
        (-\gridshadvec + \gridfadvec)\transpose]\nonumber
\\&\quad\times{\expected}\inv[(-\gridshadvec + \gridfadvec)(-\gridshadvec
+ \gridfadvec)\transpose]
(\gridpowvec-\gridbasepowvec)\nonumber\\
&= \updatelinearvec\tnot{\tind}\transpose \gridpowvec + \updatelinearoffset\tnot{\tind}
\\
\var&[\measpow\tnot{\tind}|\gridpowvec]
=
\var[\measpow\tnot{\tind}]
-
\cov[\measpow\tnot{\tind},\gridpowvec]
{\cov}\inv[\gridpowvec]
\cov[\gridpowvec,\measpow\tnot{\tind}]\nonumber\\
&=
\ushadvar
+\measandfadvar
-
(\covmat_{\shad(\loc\tnot{\tind}),\gridshadvec}
+
\covmat_{\fad(\loc\tnot{\tind}),\gridfadvec})
(\covmat_{\gridshadvec} + \fadvar \bm I_\gridnum)\inv\nonumber\\&\quad\times
(\covmat_{\shad(\loc\tnot{\tind}),\gridshadvec}
+
\covmat_{\fad(\loc\tnot{\tind}),\gridfadvec})\transpose
\define\measlikelihoodvar\tnot{\tind},
\end{align}
where the quantities
\begin{myitemize}%
\myitem\cmt{}$\updatelinearvec\tnot{\tind}\define
(\covmat_{\gridshadvec} + \fadvar \bm I_\gridnum)\inv
(\covmat_{\shad(\loc\tnot{\tind}),\gridshadvec}
+
\covmat_{\fad(\loc\tnot{\tind}),\gridfadvec})\transpose
$ and 
$ \updatelinearoffset\tnot{\tind}
\define
\basepow(\loc\tnot{\tind}) -
(\covmat_{\shad(\loc\tnot{\tind}),\gridshadvec}
+
\covmat_{\fad(\loc\tnot{\tind}),\gridfadvec})
(\covmat_{\gridshadvec} + \fadvar \bm I_\gridnum)\inv
\gridbasepowvec
$ have been defined along with
\myitem\cmt{}$\covmat_{\shad(\loc\tnot{\tind}),\gridshadvec}\define \cov[\shad(\loc\tnot{\tind}),\gridshadvec]$
and $\covmat_{\fad(\loc\tnot{\tind}),\gridfadvec}
\define\cov[\fad(\loc\tnot{\tind}),\gridfadvec]$. Clearly, the latter
contains a single non-zero entry if  $\loc\tnot{\tind}\in\grid$ and
vanishes otherwise.
\end{myitemize}%

\end{myitemize}%
\end{myitemize}%
\cmt{recombine}Finally,  \eqref{eq:bayes} and
\cite[eq. (2.116)]{bishop2006} imply that
$p(\gridpowvec|\measpowvec\tnot{\tind})=\normal(\gridpowvec|
\meanvec_{\gridpowvec|\measpowvec\tnot{\tind}},\covmat_{\gridpowvec|\measpowvec\tnot{\tind}})$
with
\begin{align}
\label{eq:onlinecovariance}
\covmat_{\gridpowvec|\measpowvec\tnot{\tind}}&=( \covmat\inv_{\gridpowvec|\measpowvec\tnot{\tind-1}}+(1/\measlikelihoodvar\tnot{\tind})\updatelinearvec\tnot{\tind}
\updatelinearvec\tnot{\tind}\transpose)\inv\nonumber\\
&=\covmat_{\gridpowvec|\measpowvec\tnot{\tind-1}}
-\frac{
 \covmat_{\gridpowvec|\measpowvec\tnot{\tind-1}}\updatelinearvec\tnot{\tind}
 \updatelinearvec\tnot{\tind}\transpose\covmat_{\gridpowvec|\measpowvec\tnot{\tind-1}}
}{
\measlikelihoodvar\tnot{\tind}+
\updatelinearvec\tnot{\tind}\transpose
\covmat_{\gridpowvec|\measpowvec\tnot{\tind-1}}
\updatelinearvec\tnot{\tind}
}
\end{align}%
and
\begin{align}
\label{eq:onlinemean}
\meanvec_{\gridpowvec|\measpowvec\tnot{\tind}}&=
\covmat_{\gridpowvec|\measpowvec\tnot{\tind}}\left[\frac{\measpow(\loc\tnot{\tind})-\updatelinearoffset\tnot{\tind}}{\measlikelihoodvar\tnot{\tind}}\updatelinearvec\tnot{\tind}
+\covmat\inv_{\gridpowvec|\measpowvec\tnot{\tind-1}}\meanvec_{\gridpowvec|\measpowvec\tnot{\tind-1}}\right].
\end{align}
The sought algorithm applies the update equations
\eqref{eq:onlinecovariance} and \eqref{eq:onlinemean} every time a new
measurement is acquired, where the initializations are given by
$\covmat_{\gridpowvec|\measpowvec\tnot{-1}}
\define\covmat_{\gridshadvec} +\fadvar\bm I_\gridnum$ and
$\meanvec_{\gridpowvec|\measpowvec\tnot{-1}}\define
\gridbasepowvec$. Unlike the batch estimator from
Sec.~\ref{subsec:batch}, Equation \eqref{eq:onlinemean} provides an
estimator
$\gridpowvecest_\tind \define
\meanvec_{\gridpowvec|\measpowvec\tnot{\tind}}$ that changes with
$\tind$, gradually accommodating new measurement information as it is
acquired.  Observe that the computational complexity is now constant
per $\tind$, and therefore this algorithm constitutes a practical
estimator for spectrum surveying.

\section{Data-Driven Uncertainty Mapping}
\label{sec:data-driven}
\cmt{Section Overview and motivation}%
\begin{myitemize}%
  \myitem\cmt{Motivation}The preceding section described techniques to estimate radio maps based on a shadowing model. Although the adopted
  formulation led to a practical online estimator, the quality of the
  map estimate may be limited if the actual propagation conditions do
  not adhere to this model. 
  \myitem\cmt{Section Overview}For this reason, this section adopts a
  data-driven perspective where the propagation phenomena are implicitly learnt from a training dataset of historical measurements collected in different
  environments. Specifically, a deep learning approach is pursued here
  to jointly obtain a radio spectral map estimate and uncertainty metric.

\end{myitemize}%

\subsection{Input and Output Representation}
\label{subsec:inputoutput}

The proposed estimator comprises a deep neural network that takes
information about the measurements at its input and returns a map
estimate and an uncertainty map at the output. This section describes
how these inputs and outputs are represented along the lines of~\cite{teganya2020rme,tang2016spectrum} to accommodate the variable number of measurements.
\cmt{Matrix
  form}
\begin{myitemize}%
  \myitem\cmt{True power}To this end, it is necessary to arrange the
  grid points as an $\griddim$ rectangular grid. By doing so, each
  grid point $\gridloc\gridnot{\gridind}$ in the notation of the
  previous sections can be identified with a row and a column in the
  rectangular grid. Thus, the grid can be written as
  $\grid = \{\gridloc\gridnot{\matind}, \matrowind=
  0,1,\ldots,\gridy-1, \matcolind=0,1,\ldots,\gridx-1\}$, where   $\gridloc\gridnot{\gridind}=\gridloc\gridnot{\matind}$ if
  $\gridind = \matrowind\gridy+\matcolind$.
  In this way, one can arrange the true received power values at
  the grid points in  matrix $\powmat \in \rfield^{\griddim}$,
  whose $(\matind)$-th entry is given by
  $[\powmat]_{\matind} = \pow(\gridloc\gridnot{\matind})$.

\nextv{To this end, one can represent the true received power values at the grid points in matrix form as $\powmat \in \rfield^{\griddim}$ whose $(\matind)$-th entry is given by $[\powmat]_{\matind} = \pow(\gridloc\gridnot{\gridind})$, where $i = \max\{p\in\mathbb{Z} \mid p\leq\gridind / \gridx\}$ and $j= \gridind- i\times\gridy$.} 
\myitem\cmt{Measurements}Similarly,  the  measurements collected until
a given time instant can be represented by a matrix of the same
dimension $\griddim$ by associating each measurement location with the
nearest grid point. 
\begin{myitemize}%
  \myitem\cmt{assign off grid measurements to grid}In particular, let
  $\setindices\gridnot{\matind}\define\{\auxtind:\|\gridloc_{\matind}-\loc_{\auxtind}\|_{2}
  \leq \|\gridloc_{\matrowind^{\prime}, \matcolind^{\prime}}
  -\loc_{\auxtind}\|_{2}$,
  $\forall \matrowind^{\prime}, \matcolind^{\prime}\}$ be a set
  containing the indices of the measurement locations assigned to
  $\gridloc_{\matind}$. The targeted matrix representation involves
  associating with $\gridloc_{\matind}$ the average
  $\powmeas\gridnot{\matind}^{(\tind)}=(1/|\setindices\gridnot{\matind}|)\sum_{\auxtind\in
    \setindices\gridnot{\matind}\cap \{0,\ldots,\tind\}}\powmeas_{\auxtind}$ if
  $|\setindices\gridnot{\matind}\cap \{0,\ldots,\tind\}|>0$ and $\powmeas\gridnot{\matind}^{(\tind)}=0$
  otherwise. 
  \myitem\cmt{Measurement matrix}The result of  aggregating all
  measurements in
  $\measpowvec_{\tind}$
  can be expressed in matrix form as
  $\powmeasmat\tnot{\tind} \in {\rfield}^{\griddim}$, where
  $[\powmeasmat\tnot{\tind}]_{\matind} = \powmeas\gridnot{\matind}^{(\tind)}$.
  \myitem\cmt{Sampling set of indices}It is also convenient to
  introduce the notation
  $\measset\tnot{\tind} = \{(\matind): |\setindices\gridnot{\matind} \cap \{0,\ldots,\tind\}|>0\} \subset
  \{0,...,\gridy-1\} \times \{0,...,\gridx-1\}$ to denote a set
  containing the indices of the grid points associated with at least
  one measurement.
  
\end{myitemize}%
\myitem\cmt{Building locations}To accommodate general scenarios, define
$\buildingset \subset \{0,...,\gridy-1\} \times \{0,...,\gridx-1\}$ to
be the set of indices of the grid points that lie inside buildings or no-fly zones.
\myitem\cmt{Mask}Furthermore, to encode the information in $\measset\tnot{\tind}$
and $\buildingset$, a mask
$\mask\tnot{\tind} \in \{1,0,-1\}^{\griddim}$ is constructed~\cite{teganya2020rme}, where $[\mask\tnot{\tind}]_{\matind} = 1$ if
$(\matind) \in \measset\tnot{\tind}$, $[\mask\tnot{\tind}]_{\matind} = -1$ if
$(\matind)\in\buildingset$, and $[\mask\tnot{\tind}]_{\matind} = 0$
otherwise. This mask is
concatenated to $\powmeasmat\tnot{\tind}$ along the third dimension to form  
\myitem\cmt{Tensor input to the n/w}$\powmeaswithmask\tnot{\tind} \in \rfield^{\griddim \times 2}$ as the tensor input to the network.

\myitem\cmt{Output}Regarding the output, functions
$\meanoutfunc: \rfield^{\griddim \times 2} \rightarrow
\rfield^{\griddim}$ and
$\stderroutfunc: \rfield^{\griddim \times 2} \rightarrow
\rfield^{\griddim}$ will be used to denote the radio map estimate and
uncertainty metric obtained by the neural network, which are thus
encoded as ${\griddim}$ matrices.
\myitem\cmt{estimate}The map estimate provided by the network given
the measurements in $\measpowvec_{\tind}$ is therefore given by
$ \gridpowvecest\tnot{\tind}=\meanoutfunc(\powmeaswithmask\tnot{\tind})$. Thus,
every time a new measurement is collected, it is necessary to update
$\powmeaswithmask\tnot{\tind}$ and evaluate the output of the network
(forward pass). 
\end{myitemize}%

\subsection{Training Loss Design}
\label{subsec:lossdesign}
\begin{myitemize}%
  \myitem\cmt{Overview}The goal now is to design a suitable training
  loss to learn the functions $\meanoutfunc$ and $\stderroutfunc$
 introduced in Sec.~\ref{subsec:inputoutput}.
 \myitem\cmt{Training dataset}Let
 $\{(\powmeaswithmask\datanot{\dataind},
 \powmat\datanot{\dataind})\}_{\dataind=0}^{\datanum-1}$ represent pairs of
 measurements and true power maps collected in $\datanum$ environments, where
 $\powmeaswithmask\datanot{\dataind}$ and $\powmat\datanot{\dataind}$
 are respectively of the form of  $\powmeaswithmask\tnot{\tind}$ and $\powmat$
 described in Sec.~\ref{subsec:inputoutput}. Subscript $\tind$ is
 omitted to simplify the notation.
 \myitem\cmt{why not krijestorac}%
 \begin{myitemize}%
   \myitem\cmt{Training with log-likelihood in the case Gaussian
     data}Given these training pairs, one could think of adopting a
   negative log-likelihood loss function to learn $\meanoutfunc$ and
   $\stderroutfunc$ as the posterior mean and posterior standard
   deviation of $\pow$ whenever the distribution of $\pow$ evaluated at an arbitrary set of locations is jointly Gaussian~\cite{krijestorac2020deeplearning}.
   \myitem\cmt{Motivation for a loss design}However, this is not
   generally the case in practice. To see this,
   Fig.~\ref{fig:histogramtruemap} depicts a histogram of $\pow(\loc)$
   for the Rossyln dataset~\cite{teganya2020rme} described in Sec.~\ref{sec:results}. It
   is observed that the distribution is markedly non-Gaussian. For
   this reason, an alternative loss function is investigated next.
\begin{figure}[t!]
    \centering
     \includegraphics[width=
     \if\dcolformat1 
        0.5\textwidth
        \else
        0.35\textwidth
     \fi]{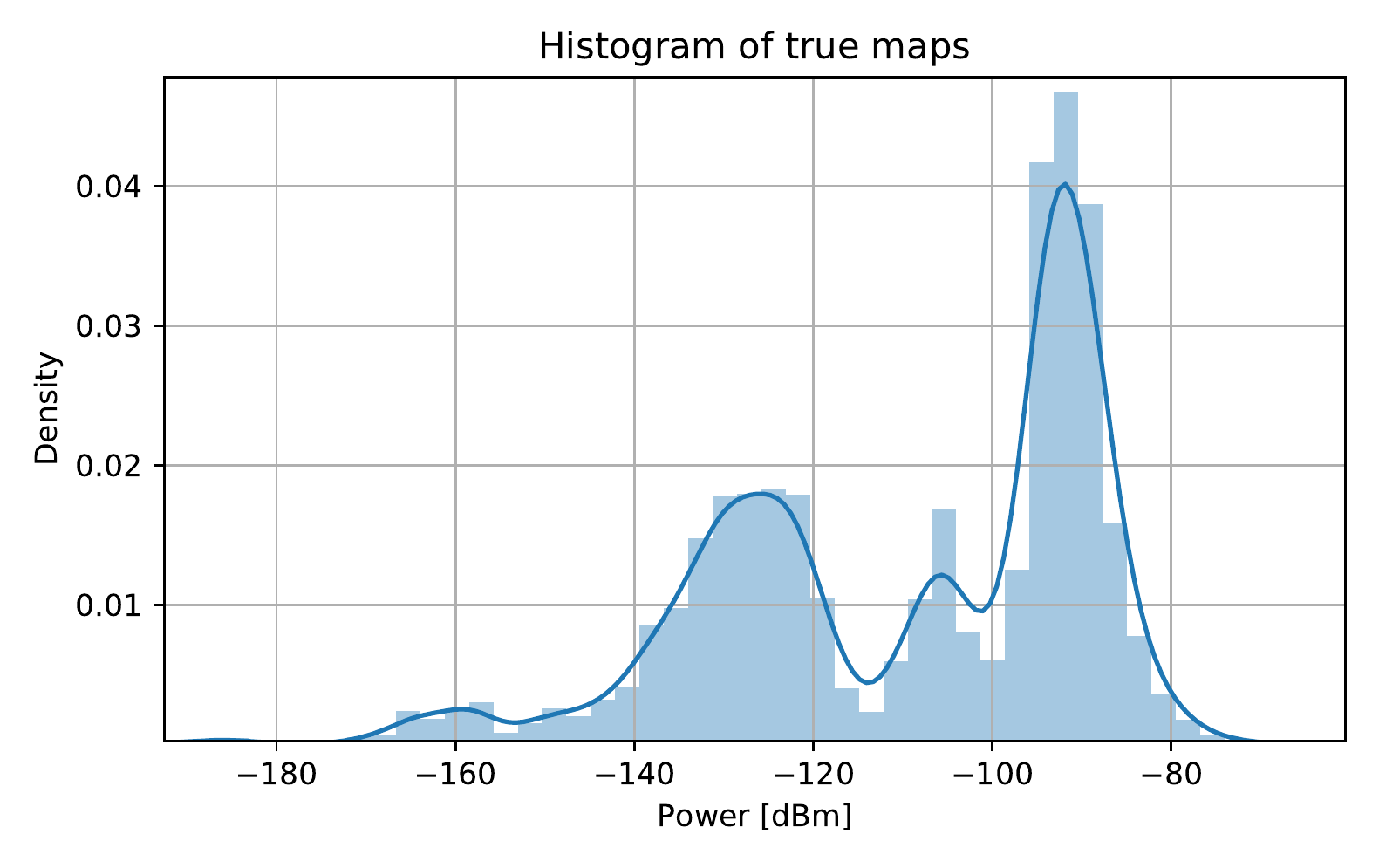}
     \caption{Histogram of true maps from the Rossyln dataset.}
    \label{fig:histogramtruemap}
  \end{figure}
\end{myitemize}
  
\myitem\cmt{Estimator for $\powmat$}
\begin{myitemize}%
  \myitem\cmt{MMSE}The idea is to rely on the well-known relation
  between the MMSE estimator and the conventional square loss. This is
  explained in detail next when designing the learning approach for
  $\meanoutfunc$ with the purpose of laying  the grounds for $\stderroutfunc$.
  Specifically, recall that the MMSE estimator of $\powmat$, i.e., the minimizer of the mean square error (MSE)
  between a target $\powmat$ and its estimate, 
  is the conditional mean
  $\expected[\powmat|\powmeaswithmask]$ \cite[Ch. 10]{kay1}
  \begin{align}
    \label{eq:mmsecondmean}
    \arg\min_{\meanoutfunc}\expected[\|\powmat-\meanoutfunc(\powmeaswithmask)\|^{2}_{F}] = \expected[\powmat|\powmeaswithmask],
\end{align}
where the right-hand side is seen as a function of $\powmeaswithmask$
and, to simplify the exposition, it is assumed that there are no
buildings in the map. The case with buildings is more involved since
measurements cannot be collected indoors. For this reason, it is
assumed that $\pow$ is not defined inside buildings. The case with
buildings is studied later.

\myitem\cmt{Loss for $\meanoutfunc$}The left-hand side of
\eqref{eq:mmsecondmean} does not directly yield an estimator for two
reasons.
\begin{myitemize}
  \myitem\cmt{expectation}On the one hand, the expectation in
  \eqref{eq:mmsecondmean} is over random pairs
  $(\powmeaswithmask,\powmat)$, but their distribution is not
  known. Thus, one can replace the expectation with a sample average,
  which  results in 
  \begin{align}
    &\arg\min_{\meanoutfunc}\expected[\|\powmat-\meanoutfunc(\powmeaswithmask)\|^{2}_{F}]\nonumber
    \\&\approx \arg\min_{\meanoutfunc}\frac{1}{\datanum}\sum_{\dataind=0}^{\datanum-1}\|\powmat\datanot{\dataind}-\meanoutfunc(\powmeaswithmask\datanot{\dataind})\|^{2}_{F}.
\end{align}
\myitem\cmt{optimization}On the other hand, the optimization in the
left-hand side of \eqref{eq:mmsecondmean} is over an arbitrary
function $\meanoutfunc$. Unfortunately, solving such a problem is only
tractable in very specific circumstances. Instead, one generally needs to
constrain the feasible set of functions e.g., to those parameterized by
a neural network. In this case, one can optimize over the set
$\{\meanfunc: \weightvec \in \rfield^P \}$ and find
\begin{align}
\label{eq:meanloss}
\estweightvec
&=\arg\min_{\weightvec}\frac{1}{\datanum}\sum_{\dataind=0}^{\datanum-1}\|\powmat\datanot{\dataind}-\meanfunc(\powmeaswithmask\datanot{\dataind})\|^{2}_{F}. 
\end{align}
Clearly, if the set $\{\meanfunc: \weightvec \in \rfield^P \}$ is
sufficiently large and $\datanum$ is sufficiently high,
then $\estmeanfunc$ will be close to the MMSE
estimator. 

\end{myitemize}%

\end{myitemize}
    
\myitem\cmt{Estimator for the Uncertainty metric}%
\begin{myitemize}%
  \myitem\cmt{MMSE for Uncertainty}As an uncertainty metric, one can
  think of estimating the posterior variance, along the lines of
  Sec.~\ref{sec:model-based}. To this end, note that the matrix
  collecting the posterior variances for all grid points is given by
  the conditional mean of the residual
  $(\powmat-\expected[\powmat|\powmeaswithmask])^{2}$, where $(.)^{2}$
  denotes entry-wise square. The key is to generalize and
  apply~\eqref{eq:mmsecondmean} in the reverse direction as before,
  which yields
\begin{align}
\text{Var}[\powmat|\powmeaswithmask] &=
                                         \expected\Big[(\powmat-\expected[\powmat|\powmeaswithmask])^{2}|\powmeaswithmask
                                         \Big]\nonumber \\&=
  \arg\min_{(\stderroutfunc)^{2}}\expected\|(\powmat-\expected[\powmat|\powmeaswithmask])^{2}-(\stderroutfunc(\powmeaswithmask))^{2}\|^{2}_{F},  
\end{align}
where $\stderroutfunc$ is chosen to denote posterior standard
deviation rather than posterior variance for reasons that will become
clear later.  \myitem\cmt{Loss for $\varmat$}If
$\expected[\powmat\datanot{\dataind}|\powmeaswithmask\datanot{\dataind}]$ were
known, one could adopt the approach in \eqref{eq:meanloss} to solve
\begin{align}
\label{eq:varloss}
\estweightvec 
=\arg\min_{\weightvec}\frac{1}{\datanum}\sum_{\dataind=0}^{\datanum-1}\dcol{\nonumber &}\|(\powmat\datanot{\dataind}-\expected[\powmat\datanot{\dataind}|\powmeaswithmask\datanot{\dataind}])^{2}\dcol{\nonumber \\ &}-(\varfunc(\powmeaswithmask\datanot{\dataind}))^{2}\|^{2}_{F}. 
\end{align}
In practice, though,
$\expected[\powmat\datanot{\dataind}|\powmeaswithmask\datanot{\dataind}]$ is
unknown. However, as argued earlier, it is close to
$\estmeanfunc$, where $\estweightvec$ is given by
\eqref{eq:meanloss}. Thus, it makes sense to replace
$\expected[\powmat\datanot{\dataind}|\powmeaswithmask\datanot{\dataind}]$ in
\eqref{eq:varloss} with
$\meanfunc(\powmeaswithmask\datanot{\dataind})$ and add the
objective in \eqref{eq:meanloss} as an additional term that promotes
that $\meanfunc$ is close to
$\expected[\powmat\datanot{\dataind}|\powmeaswithmask\datanot{\dataind}]$.

\end{myitemize}%
\myitem\cmt{Combined loss function}%
\begin{myitemize}%
\myitem\cmt{Combining both losses}This gives rise to 
\begin{align}
\label{eq:loss}
\estweightvec =& \arg\min_{\weightvec}~\Big\{(1-\lossweight)\frac{1}{\datanum}\sum\limits_{\dataind=0}^{\datanum-1} \|\powmat\datanot{\dataind}-\meanfunc(\powmeaswithmask\datanot{\dataind})\|^{2}_{F}
\nonumber
\\&
+\lossweight\frac{1}{\datanum}\sum\limits_{\dataind=0}^{\datanum-1} \|(\powmat\datanot{\dataind}- \meanfunc(\powmeaswithmask\datanot{\dataind}))^{2}-(\varfunc(\powmeaswithmask\datanot{\dataind}))^{2}\|^{2}_{F}\Big\},
\end{align} 
where 
\begin{myitemize}%
  \myitem\cmt{loss weight $\lossweight$}$\lossweight \in [0, 1]$ can
  be adjusted to balance the trade-off between  
  learning $\meanfunc$ and  $\varfunc$. 
\end{myitemize}%
\myitem\cmt{Absolute value of difference}Observe that the second term
in \eqref{eq:loss} involves squares of squares of its arguments. This
may render its scale very different from the one of the first term,
which may produce numerical problems and hinder the selection of
$\lossweight$. To alleviate such issues, $\varfunc$ is trained to
learn the absolute residual $|\powmat_{\dataind}-
\meanfunc(\powmeaswithmask\datanot{\dataind})|$ rather than the squared
residual $(\powmat\datanot{\dataind}-
\meanfunc(\powmeaswithmask\datanot{\dataind}))^2$, where $|\cdot |$
denotes entry-wise absolute value. 
This suggests training  the DNN by solving
\begin{align}
\label{eq:combinedloss}
\estweightvec = \arg\min_{\weightvec} \frac{1}{\datanum}\sum\limits_{\dataind=0}^{\datanum-1} \Big\{(1-\lossweight)\|\meanloss\|^{2}_{F} +
\lossweight\|\sigmaloss\|^{2}_{F}\Big\}, 
\end{align} 
where
\begin{myitemize}%
\myitem\cmt{}$\meanloss = \powmat\datanot{\dataind} - \meanfunc(\powmeaswithmask\datanot{\dataind})$, and
\myitem\cmt{}$\sigmaloss = |\meanloss| - \varfunc(\powmeaswithmask\datanot{\dataind})$. 
\end{myitemize}%
\end{myitemize}%

\myitem\cmt{Promoting low-uncertainty at meas. loc.}Several
improvements of \eqref{eq:combinedloss} are discussed next.%
\begin{myitemize}%
  \myitem\cmt{Motivation for impr. loss} Let us begin by noting that a good map estimate must be close to the measurements at the corresponding measurement locations, and that a reasonable uncertainty
  metric is expected to yield low values near measurement
  locations. To promote this behavior, one may  assign a
  heavier weight to the residuals at the measurement locations.
  \myitem\cmt{buildings and measurements indices sets for d-th map}Let
  set $\measset\datanot{\dataind}$ contain the indices of the measurement
  locations of the $\dataind$-th training example
  \myitem\cmt{weight scaling factor}and let $\weightscaling \in [0.5, 1]$
  represent a scaling factor.
  \myitem\cmt{loss}In this case, one may consider solving 
\begin{align}
\label{eq:improvisedloss}
\estweightvec =\arg\min_{\weightvec} \frac{1}{\datanum}\sum\limits_{\dataind=0}^{\datanum-1} \Big\{&(1-\lossweight)
\|\meanloss \odot \weightscalmat_{\dataind}\|^{2}_{F} 
\dcol{\nonumber \\
&}+
\lossweight\|\sigmaloss \odot \weightscalmat_{\dataind}\|^{2}_{F}\Big\}, 
\end{align}
where
\begin{myitemize}%
  \myitem\cmt{weight scaling
    matrix} $\weightscalmat_{\dataind} \in \rfield^{\griddim}$
  has entries
  $[\weightscalmat_{\dataind}]_{\matind} = \weightscaling$ if
  $(\matind)\in \measset\datanot{\dataind}$ and
  $[\weightscalmat_{\dataind}]_{\matind}=1-\weightscaling$ otherwise.
  \myitem\cmt{role of weight scaling factor}If $\weightscaling=0.5$,
  the minimizer of \eqref{eq:improvisedloss} boils down to the
  minimizer of \eqref{eq:combinedloss}. Setting $\weightscaling>0.5$
  increases the focus on observed locations.
\end{myitemize}%

\myitem\cmt{buildings} For scenarios with
  buildings and no-fly zones, let $\buildingset_{\dataind}$ represent the set of
  indices of the grid points inside buildings in the $\dataind$-th
  training example. Since no measurements can be acquired indoors,
  i.e. $\buildingset_{\dataind}\cap \measset\datanot{\dataind}=\emptyset$, it
  would make no sense to fit the DNN to any value there. In other words, one can assign a zero weight to the residuals inside
  buildings by setting
  $[\weightscalmat_{\dataind}]_{\matind} = \weightscaling$ if
  $(\matind)\in \measset\datanot{\dataind}$,
  $[\weightscalmat_{\dataind}]_{\matind}=0$ if
  $(\matind) \in \buildingset_{\dataind}$, and
  $[\weightscalmat_{\dataind}]_{\matind}=1-\weightscaling$ otherwise.

\end{myitemize}%
\end{myitemize}%

\subsection{Network Architecture}
\label{subsec:networkarchitecture}
\cmt{Overview}The previous section assumed a generic
family of functions
$\{(\meanfunc,\varfunc), \weightvec\in \rfield^P\}$. The form  of
these functions is next specified by developing a 
network architecture suitable for uncertainty-aware radio map
estimation. 

\cmt{Cascaded network}%
\begin{myitemize}%
  \myitem\cmt{High Level view of the network}At the high level,
  the two outputs of the DNN will be implemented here
  using a separate subnetwork with dedicated parameters. Specifically,
  let these subnetworks be denoted by $\meannn$ and $\varnn$, whose
  weights are collected respectively in $\weightvec_{\powmat}$ and
  $\weightvec_{\uncertnn}$.
  \myitem\cmt{mean subnetwork}%
  \begin{myitemize}%
    \myitem\cmt{overview}Subnetwork $\meannn$ takes $\powmeaswithmask$ as
    its input and returns an estimate of $\powmat$ as its output. The
    adopted architecture for this subnetwork is a convolutional
    autoencoder given its well-documented merits in radio map
    estimation~\cite{teganya2020rme}.
    
    \myitem\cmt{autoencoders}%
    \begin{myitemize}%
      \myitem\cmt{Definition}An autoencoder is a type of neural
      network $\meannn$ that can be expressed as \cite[Ch. 14]{goodfellow2016deep}:
        $\meannn(\powmeaswithmask)
        =\decoder_{\decoweightvec}(\encoder_{\encoweightvec}(\powmeaswithmask))$, 
      where 
      \begin{myitemize}
        \myitem\cmt{Enco. and deco.}functions $\encoder_{\encoweightvec}$ and
        $\decoder_{\decoweightvec}$ are known respectively as encoder and decoder and
        \myitem\cmt{weights of enco. and deco.}$\encoweightvec$ and
        $\decoweightvec$ denote their  associated weight vectors.
      \end{myitemize}%
      \myitem\cmt{latent variable}The output of the encoder
      $\latent = \encoder_{\encoweightvec}(\powmeaswithmask) \in
      \rfield^{N_{\latent}}$, which is known as the code or vector of
      latent variables, is of a (possibly much) smaller dimension
      $N_{\latent}$ than the dimension of $\powmeaswithmask$. Training
      the autoencoder forces the encoder to condense or summarize the
      information of $\powmeaswithmask$ into just $N_{\latent}$
      variables. 
    \myitem\cmt{Relevance of autoencoders in our
      work}This is useful to exploit the prior information that the
    input $\powmeaswithmask$ lies on a low-dimensional manifold
    embedded in a high dimensional space. This is the case of radio
    maps, as established in~\cite{teganya2020rme}. 

  \end{myitemize}%

  \myitem\cmt{convolutional}The reason for implementing an autoencoder
  using convolutional layers~\cite{ribeiro2018study} is  the ability
  of the latter 
  to efficiently learn and exploit spatial correlation, as reflected
  in their so-called \emph{translational equivariance} property. As
  detailed in~\cite{teganya2020rme}, this is highly beneficial in
  the case of radio map estimation, where the signal strength is
  substantially correlated at nearby locations. 
  \end{myitemize}%

  \myitem\cmt{uncertainty subnetwork}Having presented the
  implementation of $\meanfunc$, the next step is to implement
  $\varfunc$. To this end, it is useful to think of which information
  must be fed at the input of $\varnn$.
  \begin{myitemize}%
    \myitem\cmt{measurements}On the one hand, the information that
    $\powmeaswithmask$ conveys about the measurement locations is
    intuitively relevant to determine the uncertainty metric since
    points that lie far away from the measurement locations are
    expected to be assigned a greater uncertainty. 
    \myitem\cmt{map estimate}On the other hand, 
  the map estimate obtained by $\meannn$ may also be useful for
  uncertainty metric computation since it provides information about
  the spatial variability of the map. 
  \end{myitemize}
  \myitem\cmt{Cascading}For this reason, the subnetwork $\varnn$ makes use of both $\meannn(\powmeaswithmask)$ and $\powmeaswithmask$, which leads to a cascaded architecture where
  $\meanfunc(\powmeaswithmask) = \meannn(\powmeaswithmask)$ and
  $\varfunc(\powmeaswithmask)= \varnn(\meannn(\powmeaswithmask),
  \powmeaswithmask)$.

  Fig.~\ref{fig:networktopview} depicts the high-level architecture of
  the proposed deep convolutional neural network, which will be termed
  \textit{deep radio map and uncertainty estimator (DRUE)}
  hereafter. Subnetwork $\varnn$ is also implemented as a
  convolutional autoencoder since it is empirically observed that this
  yields a satisfactory performance; see Sec.~\ref{sec:results}. 
    \begin{figure}[t!]
      \centering
      \includegraphics[width=0.5\textwidth]{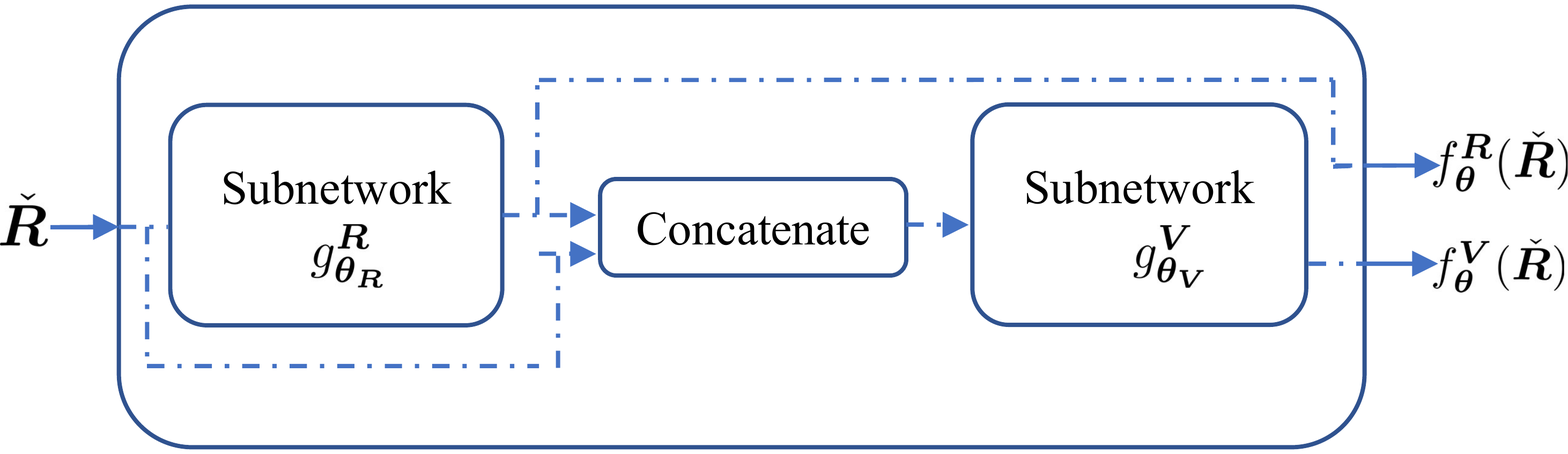}
      \caption{A high level architecture of a deep radio map and uncertainty estimator  (DRUE).}
      \label{fig:networktopview}
    \end{figure}%
\end{myitemize}%
\begin{figure}[t!]
  \centering
  \includegraphics[height=0.8\textwidth]{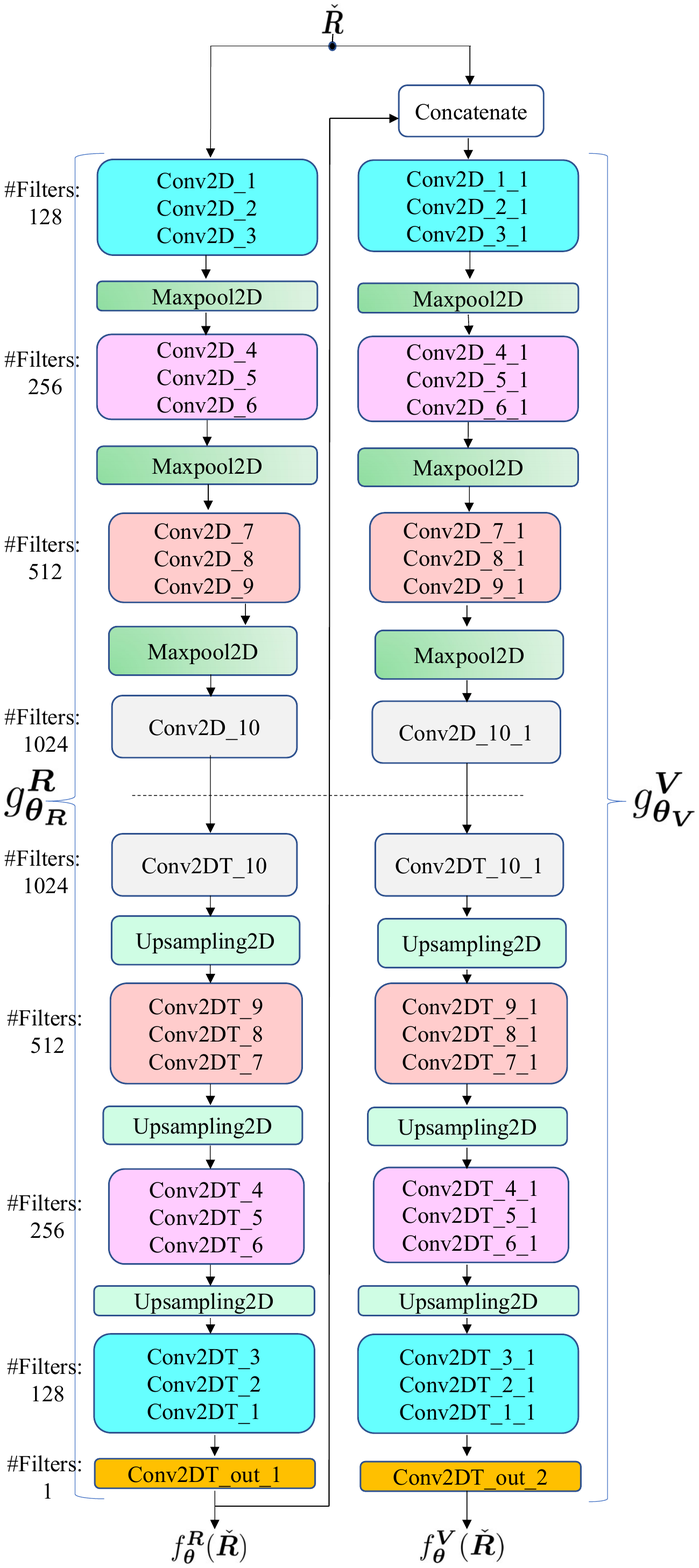}
  \caption{A detailed view of DRUE based on an autoencoder architecture. DRUE has two subnetworks $\meannn$ and $\varnn$, where each subnetwork is a fully convolutional autoencoder with an identical structure.}
  \label{fig:networkarch}
\end{figure}

\cmt{Network Parameters}%
\begin{myitemize}%
  \myitem\cmt{Detailed view}Implementation details of DRUE are shown in
  Fig.~\ref{fig:networkarch}.
  \myitem\cmt{layers}The encoder comprises mostly \textit{2D
    convolutional} and \textit{max pooling} layers whereas the decoder
  comprises \textit{2D convolutional transpose} layers, also known as
  a \textit{deconvolutional} layers, and \textit{up-sampling} layers.
  \myitem\cmt{activation func.}\textit{Leaky rectified linear units}
  (Leaky ReLUs) \cite{xu2015empirical} are utilized as activations in
  all \textit{convolutional} and \textit{deconvolutional} layers
  except in the output layer of $\varnn$, which utilizes an
  \textit{exponential} \cite{clevert2015fast} activation to guarantee
  the non-negativity of the uncertainty. Each \textit{convolutional}
  and \textit{deconvolutional} layer has a kernel size of $4 \times 4$, a stride of $1$, and same padding. \textit{Max pooling} layers have a pool size of
  2 and a stride of 2, whereas \textit{upsampling} layers have an
  upsampling factor of 2.
\end{myitemize}%

\section{Uncertainty-Aware Trajectory Planning}
\label{sec:routeplanning}
\begin{myitemize}%
  \myitem\cmt{Overview}The uncertainty metrics provided by the
  estimators in Secs.~\ref{sec:model-based} and \ref{sec:data-driven}
  are used in this section to develop a path planning scheme for
  measurement collection with a UAV.
\end{myitemize}%

\begin{myitemize}%
  \myitem\cmt{Uncertainty metric}The uncertainty in the target map at grid point
  $\gridloc_{\gridind}$ after observing $\measpowvec\tnot{\tind}$ will
  be denoted by
  $\navuncert_{\gridind}(\measpowvec_{\tind})\in \rfield_{+}$.
\begin{myitemize}%
\myitem\cmt{Two forms}%
\begin{myitemize}%
  \myitem\cmt{For Bayesian estimators}In the case of the Bayesian
  estimators from Sec.~\ref{sec:model-based},
  $\navuncert_{\gridind}(\measpowvec_{\tind})$ will be given by
\begin{align}
\label{eq:poweruncertainty}
\navuncert\gridnot{\gridind}(\measpowvec\tnot{\tind})
= \big[\covmat_{\gridpowvec|\measpowvec\tnot{\tind}}\big]_{\gridind,\gridind}
\end{align}
if there is a single transmitter. When there are multiple
transmitters, the values of the right-hand side of
\eqref{eq:poweruncertainty} for all transmitters can be averaged to
obtain a single value of $\navuncert_{\gridind}(\measpowvec\tnot{\tind})$ per
$\gridind$.
\myitem\cmt{For the DNN}For the estimator in Sec.~\ref{sec:data-driven},
$\navuncert\gridnot{\gridind}(\measpowvec\tnot{\tind})$  is given
by
\begin{align}
\label{eq:poweruncertaintynn}
\navuncert\gridnot{\gridind}(\measpowvec\tnot{\tind})
= \big[\text{vec}(\varfunc(\powmeaswithmask\tnot{\tind}))\big]\gridnot{\gridind},
\end{align}
where $\powmeaswithmask\tnot{\tind}$ is obtained from
$\measpowvec\tnot{\tind}$ as described in
Sec.~\ref{subsec:inputoutput}.  In this case, the estimator directly
provides the aggregate uncertainty for all transmitters and,
therefore, no averaging is required.
\end{myitemize}%

\myitem\cmt{Total}Finally, as a performance metric, it is convenient
to define the total uncertainty of the map after observing
$\measpowvec_{\tind}$ as the spatial average of the point-wise
uncertainty values. Since the uncertainty metric is only meaningful
outside buildings or no-fly zones (cf. Sec.~\ref{subsec:lossdesign}),
this average becomes
\begin{align}
    \label{eq:totaluncert}
    \navuncert(\measpowvec\tnot{\tind})\define
  \frac{1}{|\buildingsetcomp|}\sum_{\gridind\in \buildingsetcomp} \navuncert\gridnot{\gridind}(\measpowvec\tnot{\tind}),
\end{align}
where $\buildingsetcomp$ is the complement of the set $\buildingset$
that contains the indices of the grid points inside buildings or
no-fly zones; see Sec.~\ref{subsec:inputoutput}. 
\end{myitemize}%
\end{myitemize}%

\subsection{Trajectory Planning}
\label{subsec:trajectory}
\begin{myitemize}%
  \myitem\cmt{Goal}Ideally, one would wish to plan a trajectory for
  measurement collection such that the error in the map estimate
  decreases as quickly as possible. However, knowing the error
  requires knowing the true map, which is not possible in
  practice. Instead, one may use the uncertainty metric to
  approximate the optimal trajectory. The intuition is that its value
  is large for those candidate locations where a measurement would be
  highly informative. Collecting measurements at those locations is
  therefore expected to greatly reduce the error. This means that the
  trajectory of the UAV must pass through those locations where the
  uncertainty is high.


\myitem\cmt{Receding horizon}%
\begin{myitemize}%
  \myitem\cmt{Challenges}Finding such a trajectory is however
  challenged by the fact that the uncertainty metric changes as new
  measurements are collected and these changes cannot be generally
  predicted. Thus, the trajectory of the UAV should be computed
  on-the-fly as measurements are acquired. However, updating the
  trajectory with the reception of every new measurement may be too
  costly and may lead to erratic behavior.
  \myitem\cmt{Simplifications}%
\begin{myitemize}%
  \myitem\cmt{Receding horizon approach}A more sensible alternative is
  to adopt a \emph{receding horizon} approach that updates the trajectory
  every $\tupdate$ measurements.
\begin{myitemize}%
  \myitem\cmt{Assumption}If $\tupdate$ is sufficiently small, it makes
  sense to assume that
  $\navuncert\gridnot{\gridind}(\measpowvec\tnot{\tind})$ remains
  approximately constant between consecutive updates at all grid
  points except where a measurement is collected, say at
  $\gridloc\gridnot{\gridind_{0}}$, in which case
  $\navuncert\gridnot{\gridind_{0}}(\measpowvec\tnot{\tind})$ becomes
  0. This is because the uncertainty at a measured location is
  expected to be 0.  From \eqref{eq:totaluncert}, measuring at location
  $\loc\tnot{\auxtind}=\gridloc\gridnot{\gridind}$ yields
\begin{align}
\navuncert(\measpowvec\tnot{\tind+\tupdate})&\approx
\navuncert(\measpowvec\tnot{\tind+\tupdate-1})-\frac{\navuncert\gridnot{\gridind}(\measpowvec\tnot{\tind+\tupdate-1})}{\gridnum}\approx\ldots\nonumber
\\
                                         &\approx\navuncert(\measpowvec\tnot{\tind})-\sum_{\auxtind=\tind}^{\tind+\tupdate-1}\frac{\navuncert\gridnot{\gridind}(\measpowvec\tnot{\auxtind})}{\gridnum}.
                                           \label{eq:uncertapprox}
\end{align}%
\end{myitemize}%

\myitem\cmt{interpretation}Under this approximation, those
trajectories for which $\navuncert(\measpowvec\tnot{\tind})$ decreases
quickly are those where measurements are collected at locations
$\gridloc\gridnot{\gridind}$ with high
$\navuncert\gridnot{\gridind}(\measpowvec\tnot{\tind})$, which agrees
with the intuition presented at the beginning of this section.
\end{myitemize}%
\end{myitemize}%
\end{myitemize}%
\cmt{Approaches}%
\begin{myitemize}%
  \myitem\cmt{TSP}If $\tupdate$ were so large that it
  were possible to collect a measurement at all grid locations,
  then one could think of formulating this problem as a
  \textit{discounted reward travelling salesman problem}
  \cite{blum2007approximation}.
\begin{myitemize}%
\myitem\cmt{NP-hardness}However, this kind of problems is known to be NP-hard; see e.g.~\cite{blum2007approximation, le2009trajectory} and references therein.
\myitem\cmt{Approximation of tsp}Thus, one could think of applying for example the heuristic in \cite{blum2007approximation}.
\begin{myitemize}%
\myitem\cmt{limitations}%
\begin{myitemize}%
  \myitem\cmt{Non adaptive}Nonetheless, such a task involves high
  complexity, which would not be motivated given the approximation
  error entailed by adopting such a large $\tupdate$; recall that the
  approximation in \eqref{eq:uncertapprox} applies only for small
  $\tupdate$. This would also mean that the entire trajectory would be
  planned using the initial uncertainty metric values and therefore it
  would not be adaptive.
  \myitem\cmt{Computational
    complexity}Besides, its complexity would render this approach
  inappropriate for real-time UAV operations with limited energy and
  computational power.
\end{myitemize}%
\end{myitemize}%
\end{myitemize}%

\myitem\cmt{Proposed approach (Dynamic programming approach)} Thus, it
is preferable to adopt an alternative approach which alternatingly
updates the uncertainty metric and plans a short trajectory through
areas of high uncertainty. 
\begin{myitemize}%
  \myitem\cmt{Setting destination}To plan this trajectory, the idea
  here will be to set a destination as the grid point with the highest
  local uncertainty and then plan a route to that destination through
  points with large uncertainty. Assume without loss of generality
  that the destination and trajectory are computed at
  $\tind = \tind\tnot{\initind}$. Thus, the destination is given by
  $\loc_{\destind}\define \gridloc\gridnot{\gridind^*}$, where  
  \begin{align}
\label{eq:destination}
\gridind^*\define \arg \max_{\gridind \in \buildingsetcomp}~ \navuncert\gridnot{\gridind}(\measpowvec\tnot{\tind\tnot{\initind}}).
\end{align}

\myitem\cmt{Computing trajectory}To design the route from the current
location to $\loc_{\destind}$,
\begin{myitemize}%
  \myitem\cmt{Cost}it is necessary to define a cost function that
  promotes trajectories through high uncertainty locations. Let
  $\costfun_\tind(\loc)$ represent the cost associated with location
  $\loc$ after observing $\measpowvec\tnot{\tind}$ and defined in such
  a way that the higher the uncertainty value, the lower the cost at
  that location. Since the uncertainty values provided by the
  estimators in  Secs.~\ref{sec:model-based} and
  \ref{sec:data-driven} are associated only with
  grid points, it is natural to set the cost of a location based on
  the uncertainty of the nearest grid point. In particular, the cost
  $\costfun_\tind(\loc)$ will be set to be
  $\costfun_\tind(\loc) =
  \decfunc(\navuncert\gridnot{\gridind}(\measpowvec\tnot{\tind}))$, where
  \begin{myitemize}%
    \myitem\cmt{}$\gridind = \arg\min_{\gridind\in\buildingsetcomp}\|\loc-\gridloc\gridnot{\gridind}\|_{2}$ and 
    \myitem\cmt{}$\decfunc(.)$ is a non-negative decreasing function. 
  \end{myitemize}%
  With this definition, one could in principle set the cost of a
  trajectory 
  $\locmat=
  [\loc\tnot{\tind\tnot{\initind}},\ldots,\loc\tnot{\tind\tnot{\destind}}]$ to be
  $\sum_{\auxtind=\tind\tnot{\initind}}^{\tind\tnot{\destind}}\costfun_\tind(\loc\tnot{\auxtind})$,
  which would assign a high cost to those trajectories through low
  uncertainty areas, as desired. However, this would not account for the distance
  between measurement points. Thus, it is more appropriate to let this
  cost be
  \begin{align}
    \label{eq:costfunc}    
    \cost(\locmat)= \sum_{\auxtind=\tind\tnot{\initind}+1}^{\tind\tnot{\destind}}\int_{\loc\tnot{\auxtind-1}}^{\loc\tnot{\auxtind}} \costfun\tnot{\tind}(\loc) d\loc. 
  \end{align}

  \myitem\cmt{A path balancing the time and cost}Although this cost
  may be convenient for finding trajectories through regions of high
  uncertainty, it may well happen that the trajectory becomes too
  wiggly and, possibly, too long. This means that the values of the
  uncertainty metric used when planning the trajectory may become
  obsolete, i.e., the approximation error of \eqref{eq:uncertapprox}
  may become too large, which would lead to highly suboptimal
  trajectories. For this reason, it may be sensible to account also
  for the time it takes to reach the destination.
    \myitem\cmt{Completion time}To this end,  let $\tfun(\locmat)$ be the
  time that the UAV needs to follow a trajectory
  $\locmat $. A
  reasonable simplification is that the UAV moves at constant speed
  $\velocity$ and, therefore,
  \begin{align}
    \label{eq:timefunc}
    \tfun(\locmat)
    = \frac{1}{\velocity}\sum_{\auxtind=\tind\tnot{\initind}+1}^{\tind\tnot{\destind}}\|\loc\tnot{\auxtind}-\loc\tnot{\auxtind-1}\|_{2}.
  \end{align}
  Observe that if one minimized $\tfun(\cdot)$ rather than
  $\cost(\cdot)$, the resulting trajectory would be the shortest path
  between the current position and the destination $\loc_\text{\destind}$. 
  Thus, there clearly exists a trade-off between minimizing
  $\tfun(\cdot)$ and $\cost(\cdot)$. A ``sweet-spot'' can be found by
  properly choosing $\locmat$ to minimize their weighted sum
  \begin{align}
    \label{eq:tradeoffobj}
    \minimize_{\locmat}~~(1-\costweight)\tfun(\locmat) + \costweight\cost(\locmat)
    \\
    \text{s.t.: }~\loc\tnot{\tind\tnot{\initind}}=\loc_{\initind}, ~\loc\tnot{\tind\tnot{\destind}}=\loc_{\destind},~ \nonumber 
    \loc\tnot{\tind}\notin\buildingset^{\region}, \forall \tind, \nonumber
  \end{align}
  where 
  \begin{myitemize}%
    \myitem\cmt{A set containing building
      locations}$\buildingset^{\region}$ represents the set containing
    the building locations and no-fly zones,
    \myitem\cmt{$\costweight$}and the smaller the value of
    $\costweight\in [0, 1]$, the shorter the completion time but also
    the lower the total uncertainty of the trajectory. 
  \end{myitemize}%
  \myitem\cmt{Reformulation}Since the UAV is flying with constant velocity $\velocity$, problem \eqref{eq:tradeoffobj} reads as
  \begin{align}
    \label{eq:reformulateobj}
    \minimize_{\locmat}~\sum_{\auxtind=\tind\tnot{\initind}+1}^{\tind\tnot{\destind}}\int_{\loc\tnot{\auxtind-1}}^{\loc\tnot{\auxtind}}\Big(\frac{1}{\velocity}(1-\costweight) + \costweight\costfun_\tind(\loc)\Big)d\loc
    \\
    \text{s.t.: }~\loc\tnot{\tind\tnot{\initind}}=\loc_{\initind}, ~\loc\tnot{\tind\tnot{\destind}}=\loc_{\destind},~ \nonumber
    \loc\tnot{\tind}\notin\buildingset^{\region}, \forall \tind. \nonumber
  \end{align}
  \begin{myitemize}%

    \myitem\cmt{Integral approximation}To solve
    \eqref{eq:reformulateobj}, it is convenient to discretize the set
    of candidate measurement locations $\loc\tnot{\auxtind}$, as
    customary in contemporary route planners
    \cite{bacha2008odin,urmson2008autonomous}. Although more general
    sets can be used, for simplicity, it is assumed from now on that
    the measurement locations must lie on the grid $\grid$. In that
    case, if $\loc\tnot{\auxtind}$ and $\loc\tnot{\auxtind-1}$ are
    adjacent horizontally, vertically, or diagonally, then the
    integral term in \eqref{eq:reformulateobj} is given by
    \begin{align}
      \label{eq:edgecost}
      \int_{\loc\tnot{\auxtind-1}}^{\loc\tnot{\auxtind}}\Big(\frac{1}{\velocity}(1-\costweight) + \costweight\costfun_\tind(\loc)\Big)d\loc
      = \nonumber
      \\
      \|\loc\tnot{\auxtind}-\loc\tnot{\auxtind-1}\|_{2}\Big(\frac{1}{\velocity}(1-\costweight) + \frac{\costweight}{2}(\costfun_\tind(\loc\tnot{\auxtind})
      +\costfun_\tind(\loc\tnot{\auxtind-1}))\Big)
    \end{align}
    since, by definition, $\costfun_\tind(\cdot)$ is a piecewise
    constant function.
    
  \end{myitemize}%
  \myitem\cmt{Shortest path based trajectory}With this simplification,
  problem \eqref{eq:reformulateobj} becomes a shortest path problem
  and, therefore, can be solved with classical algorithms such as the
  well-known Bellman-Ford algorithm~\cite{bellman1958routing}. The transition costs are given by
  \eqref{eq:edgecost}.
\end{myitemize}%
\myitem\cmt{summary}To sum up, the UAV iteratively sets a destination
using \eqref{eq:destination} and plans a trajectory using
\eqref{eq:reformulateobj}.
\myitem\cmt{truncation}Observe that it is possible that the number of
measurements of the resulting trajectory is greater than $\tupdate$
since the opposite is not enforced by \eqref{eq:reformulateobj}. In
that case, the UAV may just follow the first $\tupdate$ points
indicated by the trajectory and replan afterwards. This would ensure
that the uncertainty values do not become obsolete.

\end{myitemize}%

\myitem\cmt{Smoothing of Uncertainty metric}Finally, depending on the
adopted estimator, it is possible that the uncertainty metric may
significantly change between updates. Since the
trajectory will depend on the uncertainty metric, it may be convenient
to smooth $\navuncert_{\gridind}(\measpowvec\tnot{\tind})$ to avoid
sudden changes in the trajectory. To this end, one can perform a
running average as follows:
\begin{align}
\label{eq:avgpoweruncertainty}
    \uncert\gridnot{\gridind}(\measpowvec\tnot{\tind})
= \navuncert\gridnot{\gridind}(\measpowvec\tnot{\tind})\uncertsmooth + \uncert\gridnot{\gridind}(\measpowvec\tnot{\tind-1})(1-\uncertsmooth),
\end{align}
where $\uncertsmooth\in (0, 1]$ is a smoothing factor.
\begin{myitemize}%
  \myitem\cmt{Interpretation $\uncertsmooth$}Clearly, the lower  $\uncertsmooth$, the more dependent
  on historical uncertainty is  $    \uncert\gridnot{\gridind}(\measpowvec\tnot{\tind})$.
\end{myitemize}%

\end{myitemize}%

\section{Experiments}
\label{sec:results}

\cmt{Overview}This section presents numerical results to evaluate the
performance of the proposed algorithms. 
    
\cmt{Simulation setup}%
\begin{myitemize}%
  \myitem\cmt{area}Spectrum surveying is carried out with a UAV that
  flies at a constant height to construct 2D power maps
  ($\regiondim =2$). The region of interest is discretized into an
  $\griddim=32\times32$ rectangular grid with grid point spacing of
  $3\text{ m}$ in both dimensions.
  \myitem\cmt{Data generation}Two
  datasets are considered.
\begin{myitemize}
\myitem\cmt{Two Datasets}%
\begin{myitemize}%
  \myitem\cmt{Gudmundson}In the Gudmundson dataset, the receive power
  is generated using the model in Sec.~\ref{subsec:radiomodel} using
  two transmitters  with power $\txpow=10$ dBm and height $20$
  m placed uniformly at random in $\region$. The path loss exponent is
  2, the carrier frequency is $2.4$ GHz, and antennas are
  isotropic. To focus on the effects of the shadowing, $\fadvar$ and
  $\measnoisevar$ are set to $0$. The shadowing component is generated
  with $\dist = 50$ m, $\ushadvar=10$, and $\ushadmean = 0$.
  \myitem\cmt{Remcom}The other dataset, referred to as the Rosslyn
  dataset, is the one from~\cite{teganya2020rme}, where maps are
  generated using the commercial 3D ray-tracing software Wireless
  Insite.  The so-called \emph{urban canyon} model is used with 6
  reflections and 2 diffractions.  42 large maps (40 for training and
  2 for testing) are generated in a square region of $700$ m side in
  the downtown of Rosslyn, Virginia, each one with a transmitter
  operating at $2.4$ GHz and placed at a different location. To
  generate each training or testing map, an $\griddim$ patch is
  extracted from two of these large maps at the same position and
  added.
\end{myitemize}%
\myitem\cmt{off grid data}For both datasets, measurements off the grid
are obtained via cubic spline interpolation.
\end{myitemize}%

\myitem\cmt{Network training setup}DRUE was implemented in TensorFlow
with a latent dimension of $4\times4=16$. The Adam optimization
algorithm \cite{kingma2014adam} is used for training the network with
a constant learning rate of $10^{-5}$ and a batch size of
64. Initially, DRUE is trained with $\lossweight=0.5$, then with
$\lossweight = 0$ (i.e., only layers of $\meannn$ are trainable), and
finally with $\lossweight= 1$ (i.e., only layers of $\varnn$ are
trainable). Furthermore, for each training map, the number of
measurements is drawn uniformly at random between $1$ and $100$ with
measurement locations drawn uniformly at random over the grid.

\myitem\cmt{Performance metrics}To evaluate the performance of the
considered schemes, two metrics are utilized. 
\begin{myitemize}%
\myitem\cmt{RMSE}One is the root mean square error (RMSE)
\begin{align}
    \text{RMSE}\tnot{\tind} = \sqrt{\frac{{}\expected\{\|\bm
  D_{\buildingsetcomp}(\gridpowvec - \gridpowvecest\tnot{\tind})
  \|^{2}_{2}\}}{ \trnb (\bm D_{\buildingsetcomp})}}
\end{align}%
where $\bm D_{\buildingsetcomp}$ is a diagonal matrix with ones and
zeros that sets to 0 all entries of
$\gridpowvec - \gridpowvecest\tnot{\tind}$ corresponding to grid
points inside buildings or no-fly zones and
$\gridpowvecest\tnot{\tind}$ is given in Secs.~\ref{subsection:online}
and \ref{subsec:inputoutput}. The expectation averages across maps and
measurement locations.  \myitem\cmt{Tot. rem. uncertainty}The other
metric is the total uncertainty defined in 
\eqref{eq:totaluncert}.

\end{myitemize}%
\end{myitemize}%

\subsection{Radio Map Reconstruction}
\cmt{Radio Map Reconstruction}This section corroborates the ability of
the proposed schemes to estimate radio maps. To focus on estimation
aspects, no route planning is considered in this section. Instead,
measurement locations are chosen uniformly at random throughout the
area. 
\begin{myitemize}%
  \myitem\cmt{Benchmark estimators}The proposed online Bayesian and
  data-driven DRUE algorithms are compared here with two competitors, namely
  the multikernel algorithm in \cite{bazerque2013basispursuit} and
  K-nearest neighbors (KNN)~\cite{altintas2011improving}. The multikernel algorithm is
  implemented with regularization parameter $10^{-7}$ and 15 Gaussian
  kernels with bandwidth parameter uniformly spaced between 1 and 200
  m. The KNN algorithm is implemented with $K=5$ nearest neighbors.

  \myitem\cmt{Reconstruction for
    Gudmundson}Fig.~\ref{fig:rmsegudmundson} shows the RMSE$_\tind$ vs. the
  number of measurements $\tind$ for the Gudmundson dataset.
\begin{myitemize}%
  \myitem\cmt{kriging } Batch kriging and the online Bayesian
  estimator know the mean power at each location. Recall from
  Sec.~\ref{sec:model-based} that the kriging algorithm provides MMSE
  estimates and, therefore, the minimum RMSE, which agrees with
  Fig.~\ref{fig:rmsegudmundson}.
\myitem\cmt{online}
The online Bayesian algorithm approximates the estimates of the batch
kriging algorithm by introducing the assumption that
$\measpow\tnot{\tind}$ and $\measpowvec\tnot{\tind-1}$ are
conditionally independent given $\gridpowvec$;
cf. Sec.~\ref{subsection:online}. Thus, in principle, the online
algorithm should perform worse than the batch version. The fact that
this is not the case shows that the error introduced by the
aforementioned assumption is negligible and therefore empirically
supports the usage of the online algorithm.
\begin{figure}[t!]
     \centering
     \includegraphics[width=
        \if\dcolformat1 
        0.5\textwidth
        \else
        0.35\textwidth
     \fi]{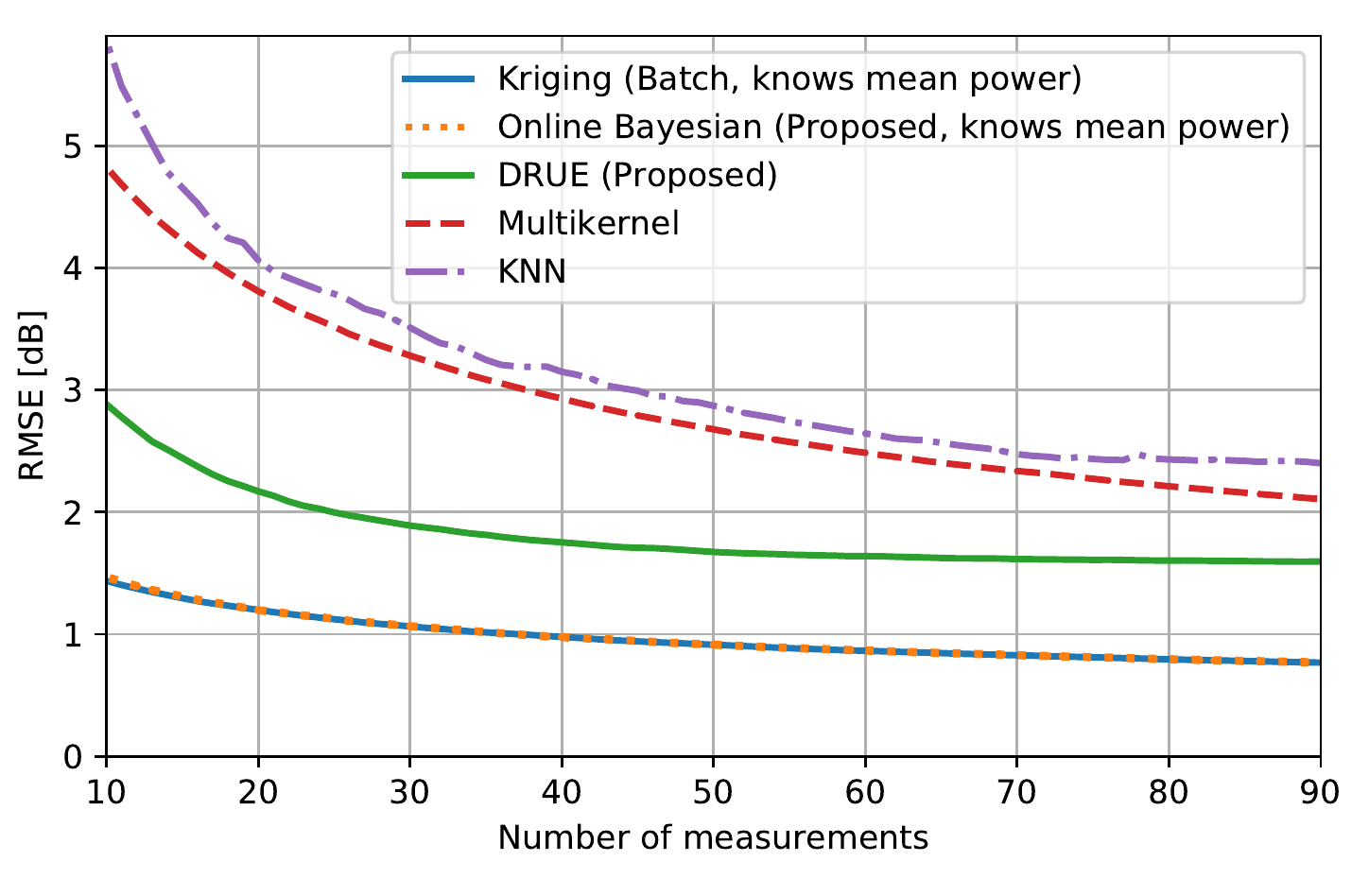}

     \caption{Comparison of the proposed estimators with the existing
       state-of-the-art for the maps generated from the Gudmundson
       dataset. The batch kriging and the proposed online Bayesian
       estimators know the mean power of the channels while the others
       do not ($\dist = 50$ m, $\ushadvar=10$, and $\ushadmean = 0$).}%
     \label{fig:rmsegudmundson}%
    \end{figure}%
\begin{figure}[t!]
     \centering
     \includegraphics[width=
     \if\dcolformat1 
        0.5\textwidth
        \else
        0.35\textwidth
     \fi]{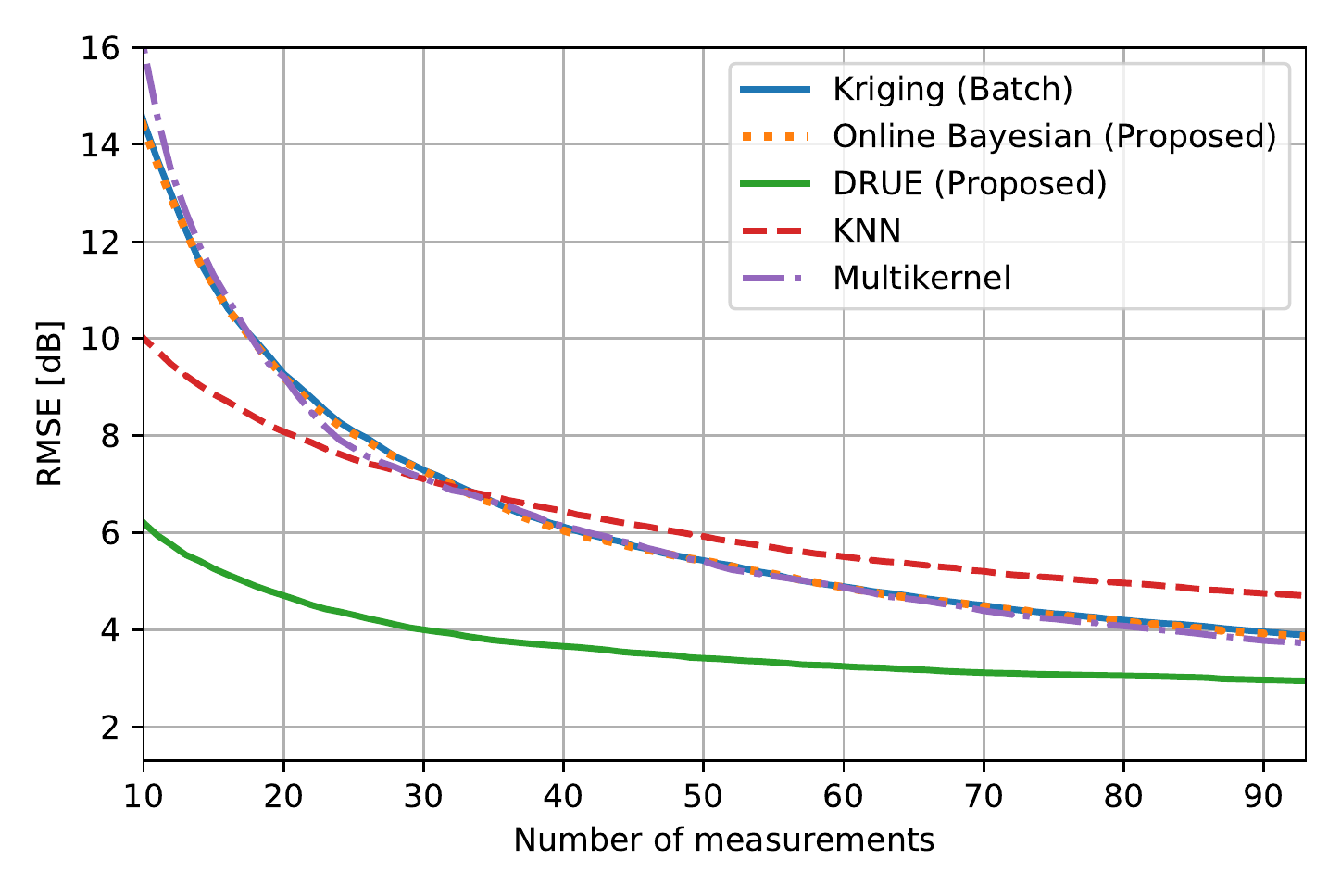}
     \caption{Comparison of the proposed estimators with the existing state-of-the-art for the maps generated from the Rossyln dataset. All the estimators do not know the mean power of the channels.}%
     \label{fig:rmseremcom}%
    \end{figure}%
    \myitem\cmt{DRUE}The performance of DRUE is slightly worse, as
    expected due to the fact that it does not know the mean. Yet, it
    outperforms the benchmarks. 

\end{myitemize}%

\myitem\cmt{Reconstruction for Rossyln}Fig.~\ref{fig:rmseremcom}
depicts the RMSE vs. the number of measurements for the Rosslyn
dataset. Two effects must be observed. 
\begin{myitemize}%
  \myitem\cmt{mean} First, since the locations and transmit power of
  the sources are unknown, the kriging and online Bayesian algorithms
  assume that the map has a zero mean. As a result, they will not
  provide (approximate) MMSE estimates.  In fact, they are
  outperformed by DRUE by a wide margin.
  \myitem\cmt{Reconst}Second, the propagation phenomena in this
  dataset is more complicated than with Gudmundson maps. The fact that
  DRUE offers the best performance corroborates its ability to learn
  from data. 
  \myitem\cmt{Online vs. Multikernel}Finally, although the multikernel
  algorithm performs similarly to the proposed online Bayesian
  estimator, observe that the former is not an online algorithm and,
  therefore, its computational complexity becomes eventually
  unaffordable if a sufficiently large number of measurements is
  collected.

\end{myitemize}%
\end{myitemize}%

\subsection{Uncertainty Learning with DRUE}
\cmt{Analysis of uncertainty}%
\begin{myitemize}%
\myitem\cmt{krijestorac2020deeplearning}%
\begin{myitemize}
  \myitem\cmt{fitness measure with a log-likelihood}If the data
  distribution were Gaussian, one could assess how well the
  uncertainty is learned by DRUE by evaluating the log-likelihood of
  test data~\cite{krijestorac2020deeplearning}.
  \myitem\cmt{Not suitable in our case}However, this is not applicable
  for the Rosslyn dataset as data there is not Gaussian;
  cf. Sec.~\ref{subsec:lossdesign}.
\end{myitemize}%
\myitem\cmt{Alternative approach}Therefore, an alternative approach is
explored here. The idea is to plot a histogram of the normalized
residuals at locations with no associated
measurements. Specifically, the normalized residual
$\pow\gridnot{\gridind}^{\prime}$ at an unobserved location
$\gridloc\gridnot{\gridind}$ is given by
\begin{align}
\label{eq:normalizedpower}
   \pow\gridnot{\gridind}^{\prime}=\frac{[\gridpowvec]\gridnot{\gridind}-[\gridpowvecest\tnot{\tind}]\gridnot{\gridind}}{[\text{vec}(\varfunc(\powmeaswithmask\tnot{\tind}))]_{\gridnot{\gridind}}},
\end{align}
where $\tind$ is fixed and $\gridind\notin \buildingset\cup\measset$. 
\begin{figure}[t!]
    \centering
     \includegraphics[width=
     \if\dcolformat1 
        0.5\textwidth
        \else
        0.35\textwidth
     \fi]{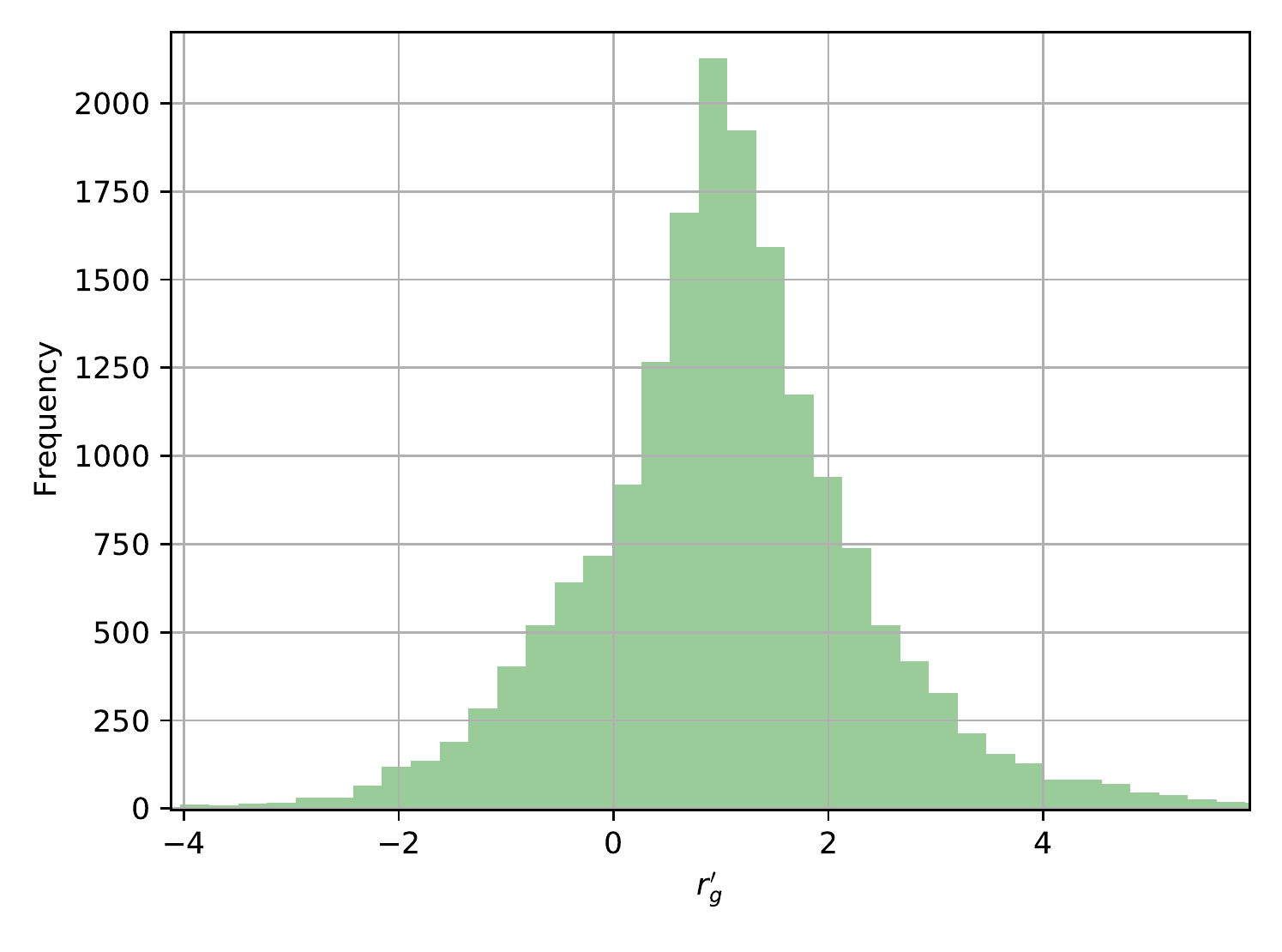}
     \caption{Histogram of the normalized residual with the Rosslyn
       dataset, where estimates of the power map and uncertainty
       metric are obtained by DRUE with $|\measset|=30$ locations
       drawn uniformly at random.}%
    \label{fig:histogramnormalizedremcom}%
    \end{figure}%
    Recall that the network is trained so that the denominator in
    \eqref{eq:normalizedpower} captures the magnitude of the error,
    which is given by the numerator. The histogram plot of
    $\pow\gridnot{\gridind}^{\prime}$ is therefore expected to be
    centered near $1$ if the uncertainty metric is estimated
    well. This is seen to be the case in
    Fig.~\ref{fig:histogramnormalizedremcom}, which shows the
    normalized histogram of $\pow\gridnot{\gridind}^{\prime}$ for the
    Rosslyn dataset, thereby confirming that the uncertainty is
    learned satisfactorily by DRUE.
\end{myitemize}%

\subsection{Spectrum Surveying Experiments}

\cmt{overview}This section illustrates the operation of the proposed
spectrum surveying scheme and compares its performance against three
other algorithms.

\cmt{Route Planner setup}The UAV is allowed to move on a horizontal
plane in one out of 8 directions that differ by $45$ degrees: East,
West, North, South, SouthEast, SouthWest, NorthEast, NorthWest.  To
enable a fair comparison across route planners, the UAV collects
measurements every $7$ m on its trajectory instead of at the
waypoints.
\cmt{Description of the estimator and cost setup}Throughout, DRUE is
used to estimate the uncertainty metric for the proposed minimum cost
planner. The reciprocal decreasing function
$\decfunc(\uncert\gridnot{\gridind}(\measpowvec\tnot{\tind})) =
1/(\uncert\gridnot{\gridind}(\measpowvec\tnot{\tind})+\posconst)$ is
used to obtain the cost function, where $\epsilon$ is a small positive
constant, and $\costweight$ is set to $0.75$.

\cmt{Map estimate with min. cost. planner} Fig.~\ref{fig:mapestimate}
shows sample trajectories using the proposed uncertainty-aware minimum
cost planner with DRUE for the Gudmundson
(Fig.~\ref{fig:mapestimate}a) and Rosslyn
(Fig.~\ref{fig:mapestimate}b) datasets. The first and second plots in
each figure
respectively correspond to the true and estimated power maps. The
third plot depicts the uncertainty metric
 after measurements are collected at the
locations indicated by red markers. The trajectory indicated by white
markers and the destination indicated by a magenta marker are computed
using this uncertainty metric. As expected, it is observed that the
resulting trajectory traverses points of high uncertainty.

\if\dcolformat1 
        \begin{figure*}[t!]
    \centering
    \begin{subfigure}[b]{1\textwidth}
     \centering
     \includegraphics[width=1\textwidth]{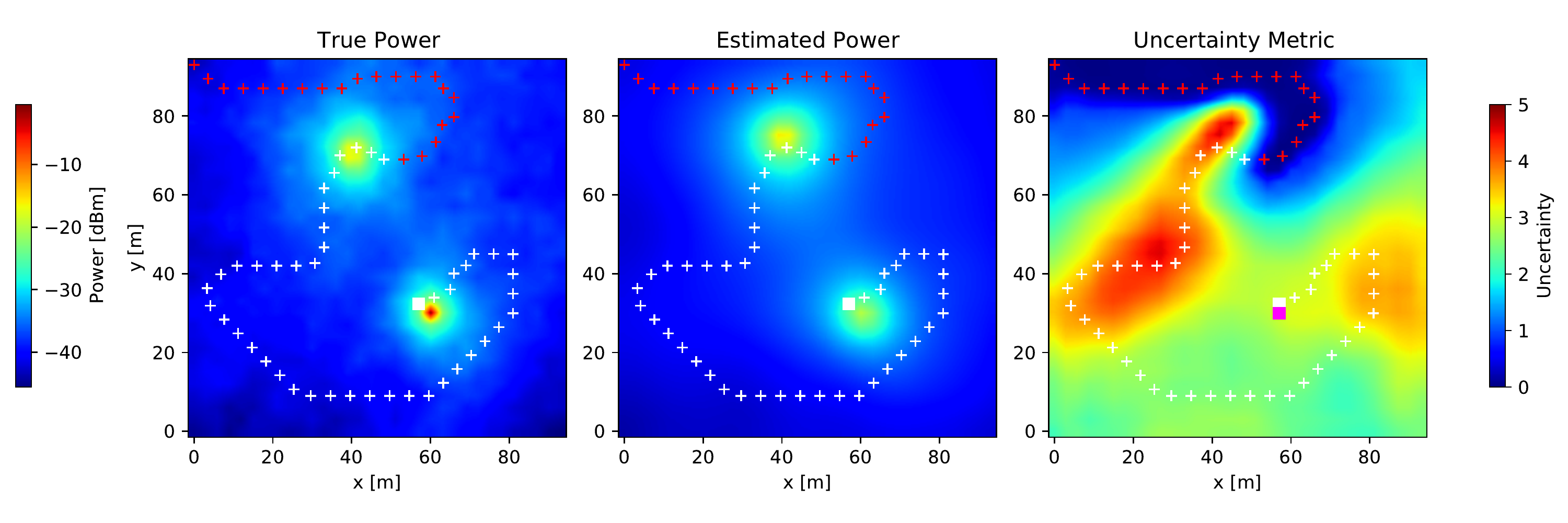}
     \caption{}
     \end{subfigure}
    \begin{subfigure}[b]{1\textwidth}
     \centering
     \includegraphics[width=1\textwidth]{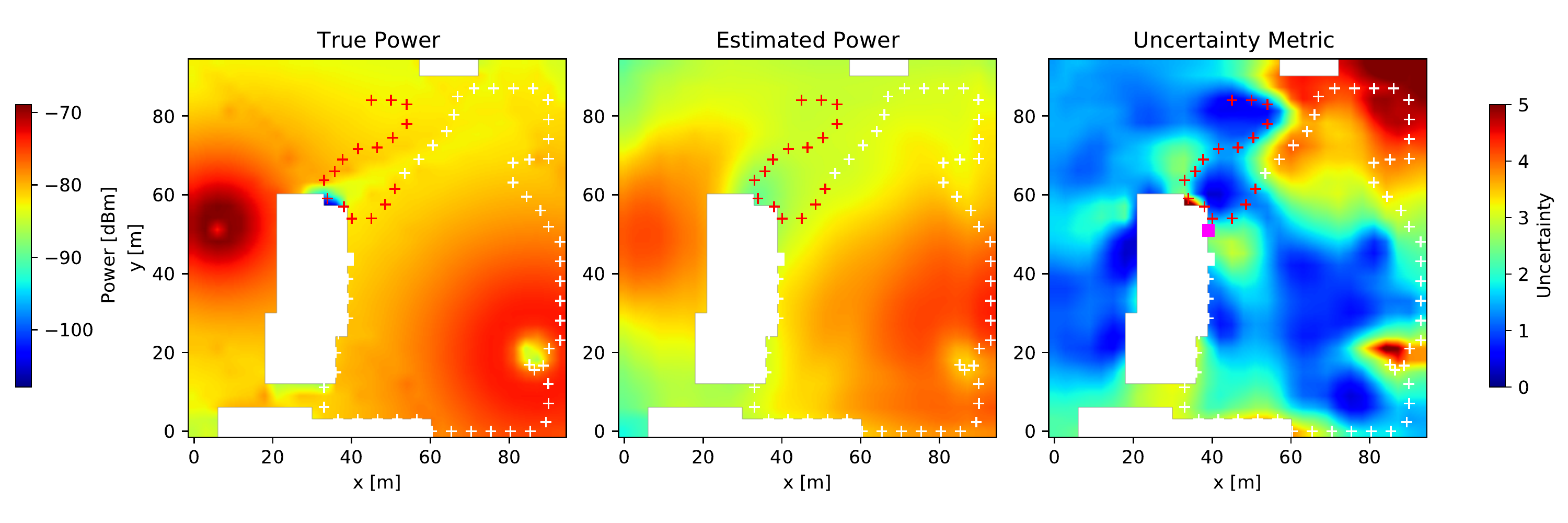}
     \caption{}
     \end{subfigure}
     \caption{Radio map estimate with the proposed uncertainty-aware minimum
       cost planner for a) the Gudmundson dataset ($\dist = 50$ m,
       $\ushadvar=10$, and $\ushadmean = 0$); and b) the Rossyln dataset
       using DRUE ( $\costweight = 0.75$, $\uncertsmooth=0.25$, and
       $\decfunc(\uncert\gridnot{\gridind}(\measpowvec\tnot{\tind})) =
       1/(\uncert\gridnot{\gridind}(\measpowvec\tnot{\tind})+\posconst)$). 
       White boxes in (b) represent buildings.}%
       \label{fig:mapestimate}
    \end{figure*}%
        \else

\begin{figure}[t!]
    \centering
    \begin{subfigure}[b]{0.49\textwidth}
     \centering
     \includegraphics[width=1\textwidth]{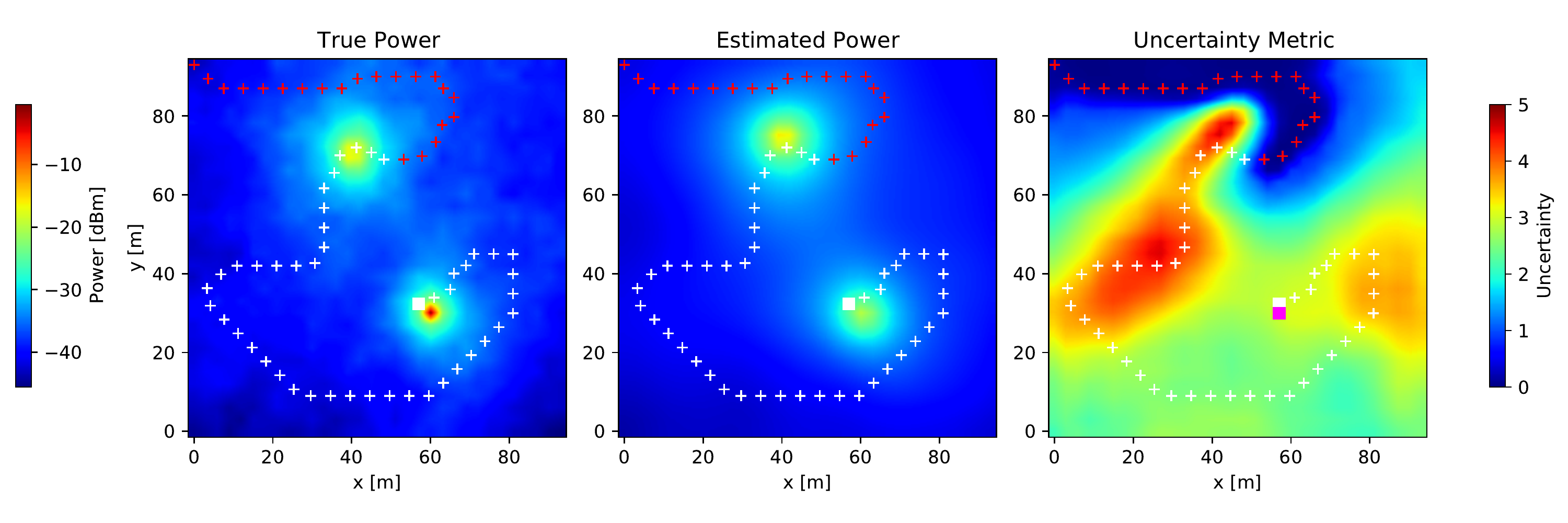}
     \caption{}
     \end{subfigure}
    \begin{subfigure}[b]{0.49\textwidth}
     \centering
     \includegraphics[width=1\textwidth]{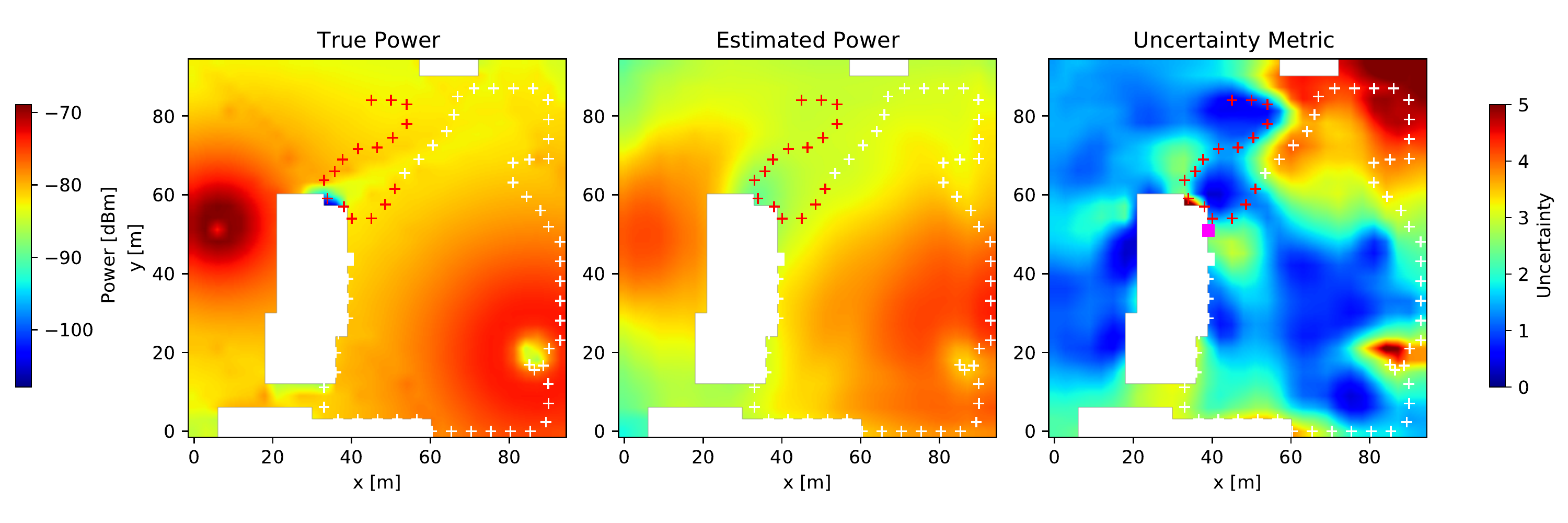}
     \caption{}
     \end{subfigure}
     \caption{Radio map estimate with the proposed uncertainty-aware minimum
       cost planner for a) the Gudmundson dataset ($\dist = 50$ m,
       $\ushadvar=10$, and $\ushadmean = 0$); and b) the Rossyln dataset
       using DRUE ( $\costweight = 0.75$, $\uncertsmooth=0.25$, and
       $\decfunc(\uncert\gridnot{\gridind}(\measpowvec\tnot{\tind})) =
       1/(\uncert\gridnot{\gridind}(\measpowvec\tnot{\tind})+\posconst)$). 
       White boxes in (b) represent buildings.}%
       \label{fig:mapestimate}
    \end{figure}%
\fi

\cmt{comparison}

\begin{myitemize}%
  \myitem\cmt{overview}The rest of this section compares the proposed
  scheme against three other algorithms. Since this is the first work
  to address \emph{adaptive} measurement collection, there are no
  competing algorithms to compare with. Instead, the proposed
  uncertainty-aware trajectory planning algorithm is compared against
  the non-adaptive approach in~\cite{zhang2020spectrum} and two more
  benchmarks.
  %
  \myitem\cmt{description of route planners}Specifically, the considered benchmarks
  are the following:
  \begin{myitemize}
    \myitem\cmt{Grid Planner}i) The \emph{grid planner} from
    \cite{zhang2020spectrum}, which traverses the grid points by
    moving column by column.  \myitem\cmt{Spiral grid Planner}ii) A
    \emph{spiral grid planner}, which starts in the top-left corner
    and moves in a rectangular spiral fashion towards the center of
    the grid.  \myitem\cmt{Indp. uniform planner}iii) An
    \emph{independent uniform planner}, which randomly selects a grid
    point and proceeds towards there in a straight line. In all cases,
    obstacles are avoided by the trajectory. 
  
    \myitem\cmt{Route planner performance
      analysis}Fig.~\ref{fig:routeremcom} shows  RMSE$_\tind$ and the total uncertainty~\eqref{eq:totaluncert} obtained via Monte
    Carlo simulation for the aforementioned route planners with DRUE
    using the Rosslyn dataset, which serves as a proxy for real-world
    propagation circumstances.
\end{myitemize}%
    \begin{figure}[t!]
    \centering
    \begin{subfigure}[b]{
    \if\dcolformat1 
        0.5\textwidth
        \else
        0.35\textwidth
     \fi}
     \centering
     \includegraphics[width=1\textwidth]{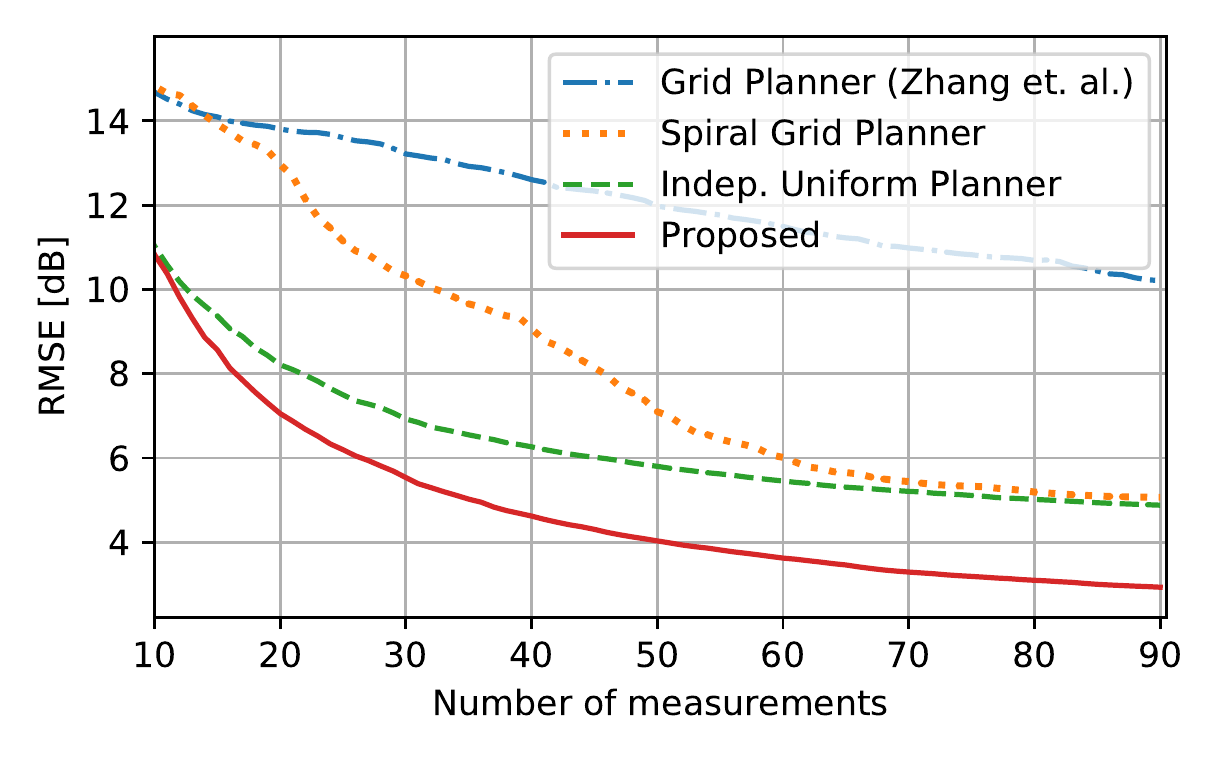}
     \end{subfigure}
    \begin{subfigure}[b]{\if\dcolformat1 
        0.5\textwidth
        \else
        0.35\textwidth
     \fi}
     \centering
     \includegraphics[width=1\textwidth]{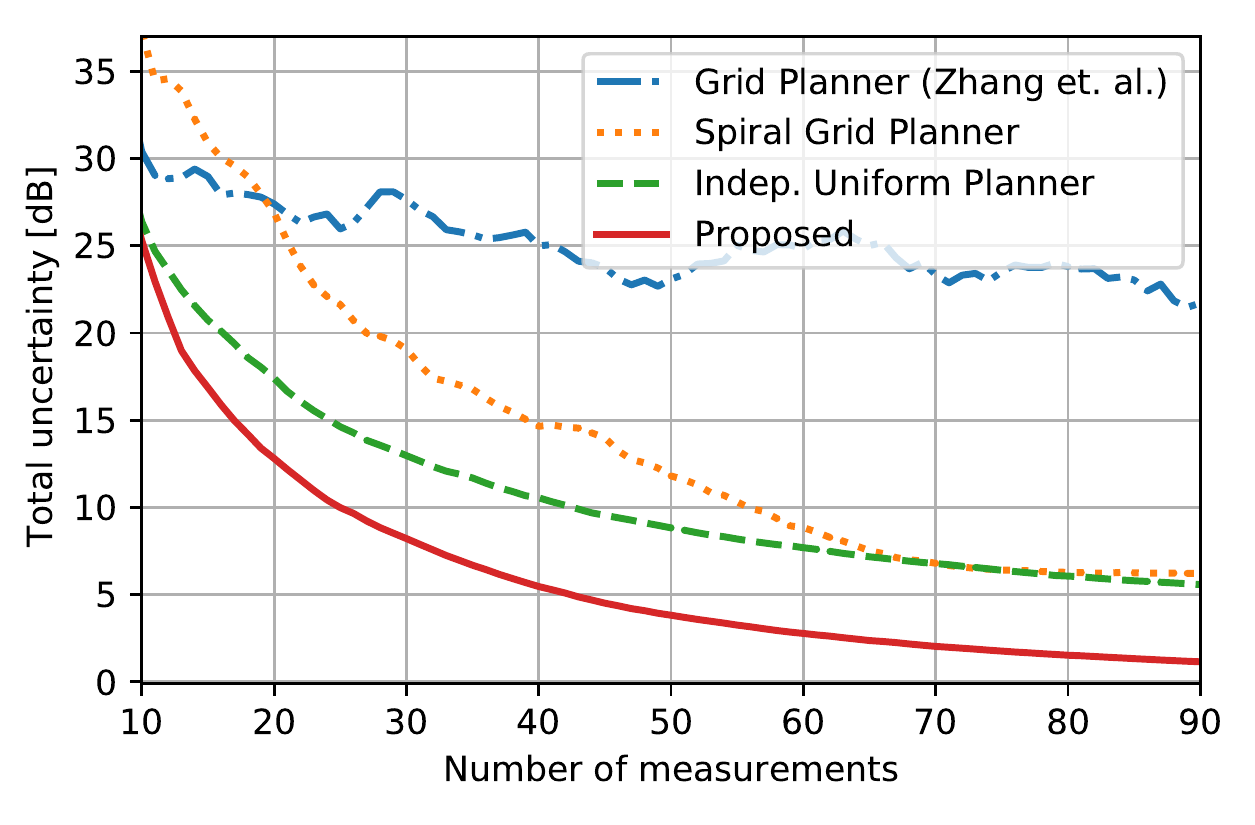}
     \end{subfigure}
       \caption{Comparison of the proposed uncertainty-aware minimum cost route planner with other three benchmarks for the Rossyln dataset using DRUE; $\costweight = 0.75$, $\uncertsmooth=0.25$, $\tupdate=7$, and $\decfunc(\uncert\gridnot{\gridind}(\measpowvec\tnot{\tind})) = 1/(\uncert\gridnot{\gridind}(\measpowvec\tnot{\tind})+\posconst)$.}%
       \label{fig:routeremcom}%
    \end{figure}%
\begin{myitemize}%
  \myitem\cmt{Proposed approach allows to construct map quickly}It is
  seen that the devised approach reduces the
  uncertainty  and error significantly faster than the benchmarks.
  \myitem\cmt{Performance in contrast to spiral and
    ind. uni. planners}In fact, the proposed algorithm requires less
  than 50\% of the measurements required by the best benchmarks to
  reach an RMSE of approximately 5 dB.
  \myitem\cmt{uncert. metric}It is also seen that the reduction in the
  uncertainty metric parallels the reduction in the error, which
  further supports that the  metric in Sec.~\ref{subsec:lossdesign}
  constitutes a reasonable uncertainty metric.
  \myitem\cmt{why grid planner performs worst}The grid planner 
  performs worst in this experiment because,  with the small number of
  measurements considered, the UAV has only time  to explore
  one side of the map. 
\end{myitemize}%
\end{myitemize}%

\subsection{Route Planner Parameters}
\label{subsec:costtrade-off}
\begin{figure}[t!]
    \centering
    \begin{subfigure}[b]{\if\dcolformat1 
        0.5\textwidth
        \else
        0.35\textwidth
     \fi}
     \centering
     \includegraphics[width=1\textwidth]{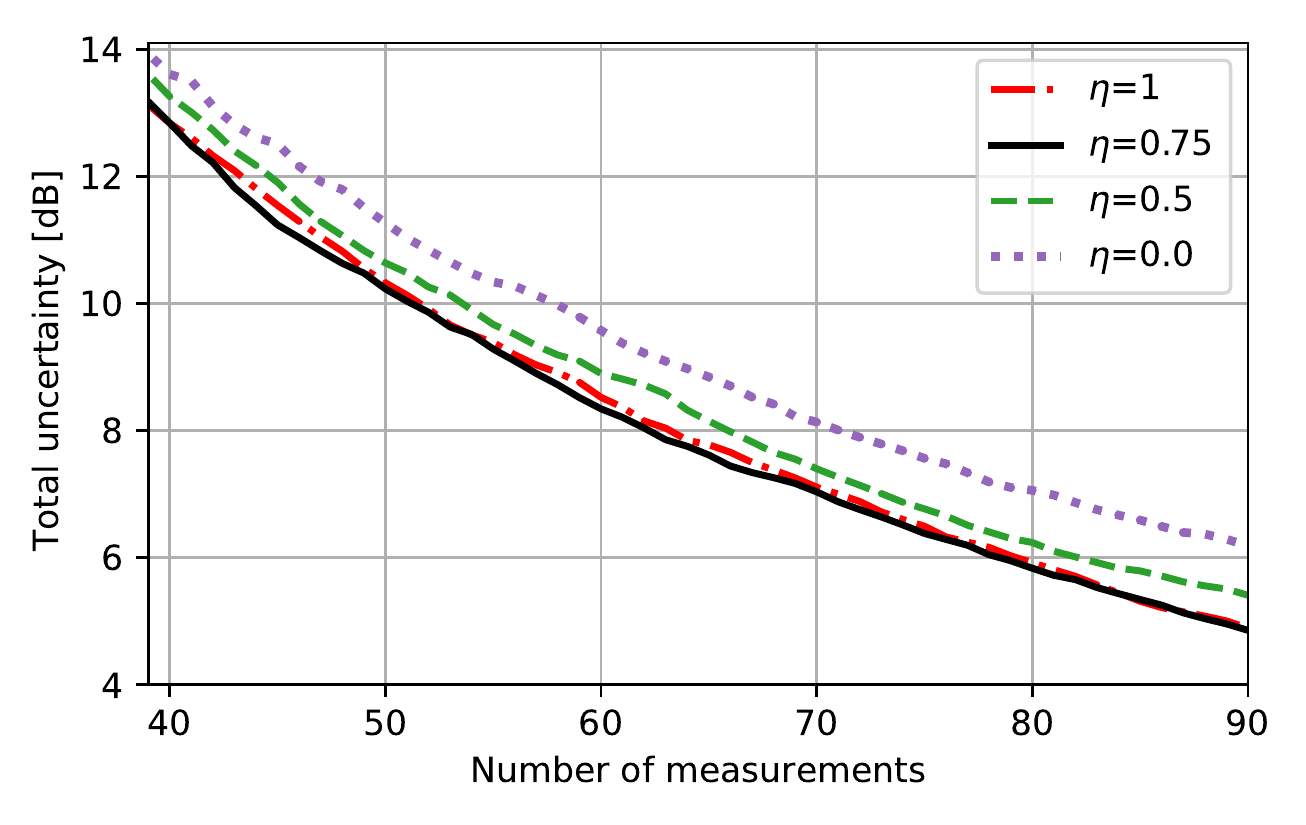}
     \end{subfigure}
       \caption{Uncertainty decrements in the map from the Rossyln dataset for different values of $\costweight$ with $\decfunc(\uncert\gridnot{\gridind}(\measpowvec\tnot{\tind})) = 1/(\uncert\gridnot{\gridind}(\measpowvec\tnot{\tind})+\posconst)$ using DRUE. The selection of $\costweight$ trade-off between the cost and the distance for the trajectory planning; $\tupdate=7$, and $\uncertsmooth=0.25$.}%
       \label{fig:costfunctiontradeoff}%
    \end{figure}

This section discusses two experiments intended to determine the best
configuration for the proposed route planner.
\cmt{Selection of decreasing function}The first experiment explores
the impact of the selection of the non-negative decreasing function
$\decfunc$ (see Sec.~\ref{subsec:trajectory}) on the uncertainty
reduction.
\begin{figure}[t!]
    \centering
    \begin{subfigure}[b]{\if\dcolformat1 
        0.5\textwidth
        \else
        0.35\textwidth
     \fi}
     \centering
     \includegraphics[width=1\textwidth]{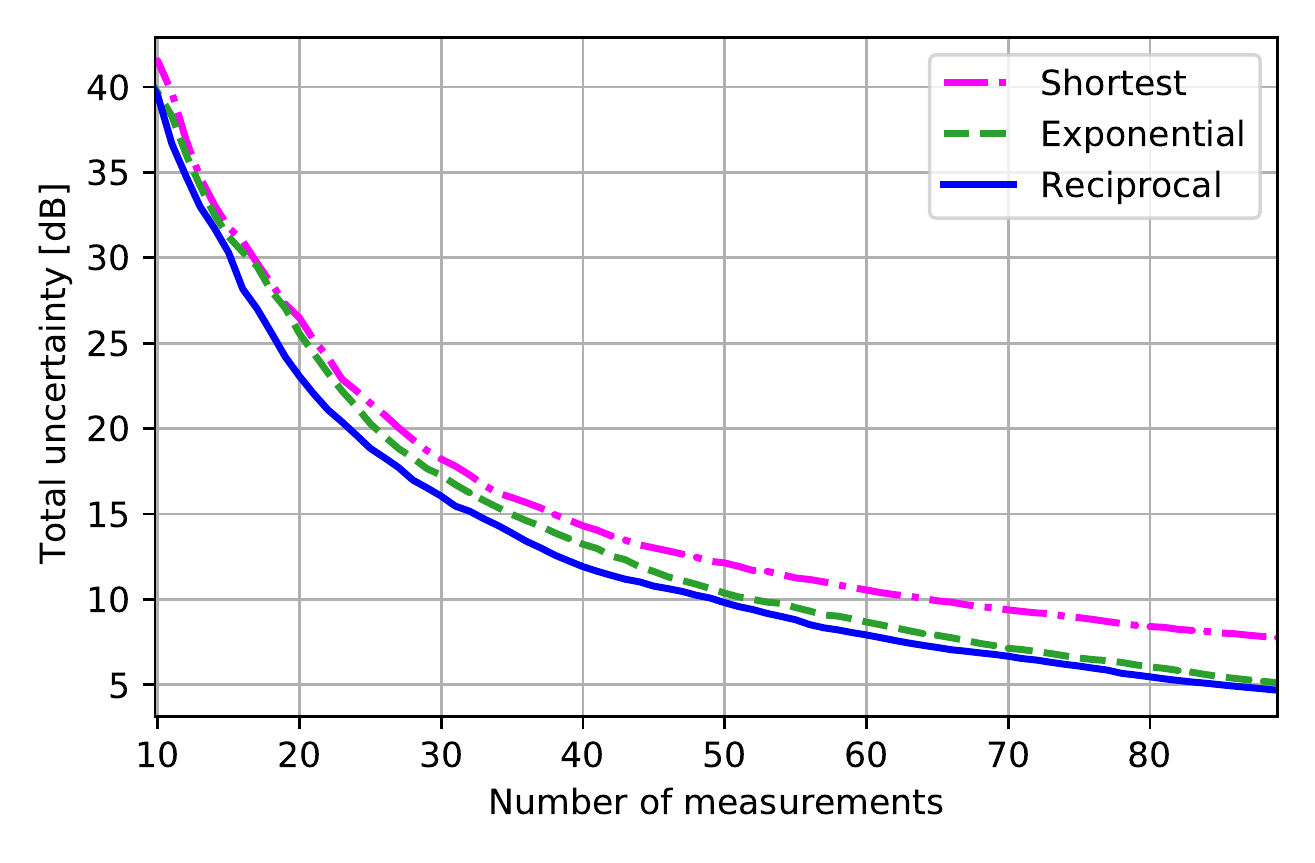}
     \label{fig:costfunctionrmseremcom}
     \end{subfigure}
    \begin{subfigure}[b]{\if\dcolformat1 
        0.5\textwidth
        \else
        0.35\textwidth
     \fi}
          \centering
     \includegraphics[width=1\textwidth]{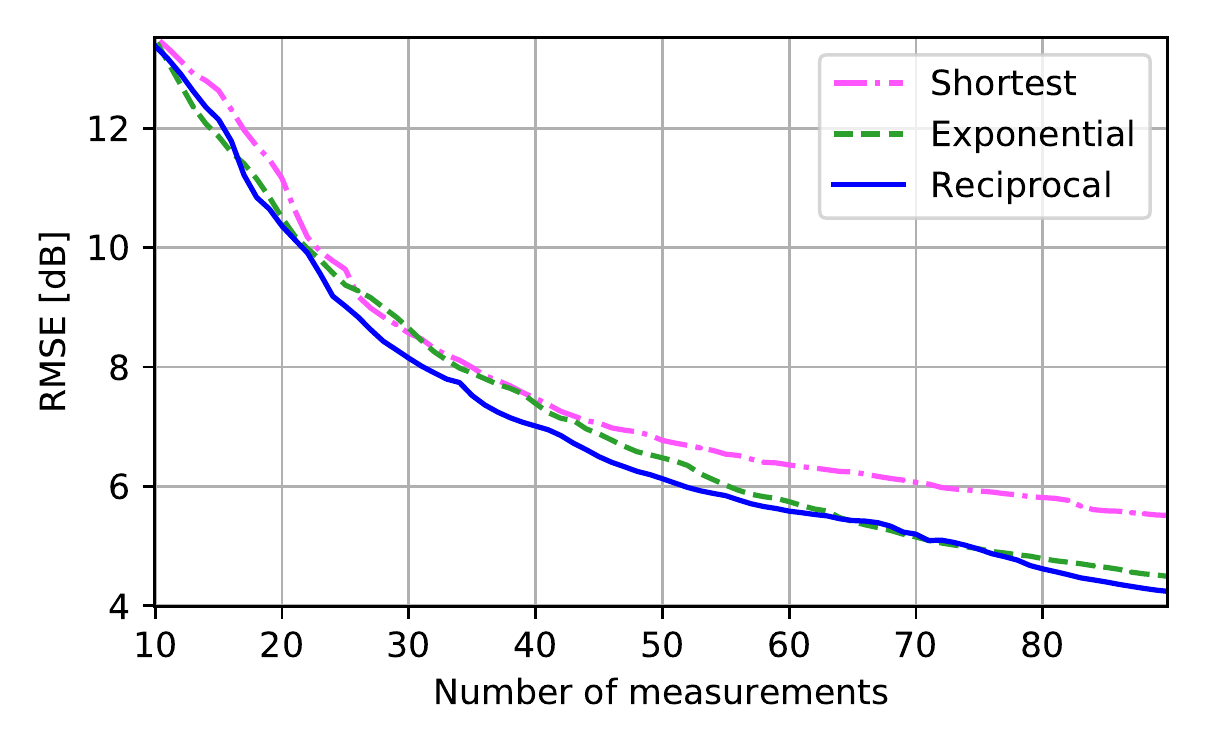}
          \label{fig:costfunctionuncertremcom}
     \end{subfigure}
       \caption{Comparison of two different cost functions with the shortest path cost function for the Rossyln dataset using the proposed route planner, and DRUE; $\costweight$ = 1, $\tupdate=7$, and $\uncertsmooth=0.25$.}%
       \label{fig:costfunctionremcom}%
    \end{figure}%
\begin{myitemize}%
  \myitem\cmt{Uncertainty decrements for different dec. func.}The
  reciprocal function
  $\decfunc(\uncert\gridnot{\gridind}(\measpowvec\tnot{\tind})) =
  1/(\uncert\gridnot{\gridind}(\measpowvec\tnot{\tind})+\posconst)$
  and the exponential function
  $\decfunc(\uncert\gridnot{\gridind}(\measpowvec\tnot{\tind}))=
  \text{exp}(-\uncert\gridnot{\gridind}(\measpowvec\tnot{\tind}))$ are
   compared with the shortest path planner, which
  results from setting
  $\decfunc(\uncert\gridnot{\gridind}(\measpowvec\tnot{\tind})) =
  1$. From Fig.~\ref{fig:costfunctionremcom}, the steepest uncertainty
  and RMSE reduction are obtained with the reciprocal function. 

  \myitem\cmt{trade-off bet. time and
    $\costweight$}Fig.~\ref{fig:costfunctiontradeoff} investigates the
  trade-off between minimizing distance and maximizing uncertainty
  collection, which is controlled by $\costweight$;
  cf. Sec.~\ref{subsec:trajectory}.  When $\costweight = 0$, the UAV follows the
  shortest path without considering uncertainty.  From
  Fig.~\ref{fig:costfunctiontradeoff}, this yields the worst
  performance. In contrast, when $\costweight = 1$, the focus is on
  traversing areas with high uncertainty. This does not result in the
  fastest uncertainty reduction either. The ``sweet spot'' is seen to be
  attained for an intermediate value of $\costweight$, namely
  $\costweight=0.75$. 
\end{myitemize}%
    
\section{Related Works}
\label{sec:relatedwork}
\cmt{Overview}This section describes the connections between the present
paper and the closely related works. 

\begin{myitemize}%
  \myitem\cmt{Krijestorac paper}In \cite{krijestorac2020deeplearning},
  the received power at every location is modeled as an independent
  Gaussian random variable and a DNN is used to learn its parameters.
  The variance estimate that it provides could be used as an
  uncertainty metric. In contrast, DRUE provides an uncertainty metric
  applicable even when the data is not Gaussian; cf.~Fig.~\ref{fig:histogramtruemap}.
  \myitem\cmt{zhang2020spectrum}The work in \cite{zhang2020spectrum}
  utilized moving sensors to collect measurements on a path specified
  by a system designer to estimate a radio map. Thus, as opposed to
  \cite{zhang2020spectrum}, the route planner proposed here is
  adaptive to measurements and radio map estimates.
  \myitem\cmt{zeng2021simultaneous}The work in
  \cite{zeng2021simultaneous} constructs a radio map of the outage
  probability using UAVs.  However, the trajectory is not planned for
  sensing purposes. Instead, the goal is to minimize the integral of
  the outage probability.  Furthermore, a fixed destination is
  necessarily given rather than updated based on the measurements
  gathered so far.

  \myitem\cmt{Mobile sensing literature}Related work in the mobile
  sensing literature includes \cite{ma2017informative,
    ling2016gaussian, popovic2017online}, which addressed the problem
  of deciding the path of a mobile robot to maximize information
  collection about a spatial field while operating under a certain set
  of path constraints.
\begin{myitemize}%
\myitem\cmt{ma2017informative}In \cite{ma2017informative}, a spatial
field is modeled as a Gaussian Process
(GP)~\cite{rasmussen2003gaussian} and the measurement locations are
selected based on the estimates of the parameters rather than directly
on the measurements.
\myitem\cmt{ling2016gaussian}The uncertainty metric proposed in
\cite{ling2016gaussian} is a sum of a function of measurement values
and a function of measurement locations. This is not general enough to
accommodate the case where the reward is given by the reduction of
uncertainty in radio map estimation since the uncertainty is a complicated
function of both measurement values and measurement locations.
\myitem\cmt{popovic2017online}The work in \cite{popovic2017online}
selects sampling locations based on whether a time-varying parameter
is less than  some random number, rather than on the available
measurements and estimates.
\end{myitemize}%
Thus, none of these schemes are directly applicable to the spectrum
surveying problem. 

\end{myitemize}%

\section{Conclusions}
\label{sec:conclusions}
\cmt{Spectrum Surveying}This paper proposed a spectrum surveying
scheme where autonomous UAVs collect radio measurements to construct a
power map. The task of spectrum surveying was decomposed into two
steps: uncertainty-aware radio map estimation and trajectory planning.
\begin{myitemize}%
\myitem\cmt{Estimation}Two estimators with complementary benefits were
developed. 
\begin{myitemize}%
  \myitem\cmt{Model-based}The first is an online Bayesian
  algorithm built upon the well-known Gudmundson shadowing model.  The
  price to be paid for its simplicity is its inability to capture
  complex propagation phenomena. 
  \myitem\cmt{DNN}To bypass this limitation, a data-driven uncertainty
  mapping technique was proposed. A tailor-made DNN architecture was
  designed to estimate power maps as well as the associated
  uncertainty even for non-Gaussian data. 
\end{myitemize}%
\myitem\cmt{Route planning}Using the uncertainty provided by these
estimators, a route planning algorithm was devised to determine
trajectories that pass through the most informative  locations.
\end{myitemize}%
\cmt{future work}As future work, one could use a variational
autoencoder to learn the posterior distribution of the radio map and
develop corresponding uncertainty metrics. 

\printmybibliography
\end{document}